\documentclass[a4paper,11pt]{article}
\pdfoutput=1 

\usepackage{graphicx}
\usepackage{amsfonts}
\usepackage{latexsym}
\usepackage{amsmath}
\usepackage{amssymb}
\usepackage{epsfig}

\usepackage{jheppub}
\usepackage{slashed}
\usepackage{multirow}
\usepackage{color}
\usepackage[normalem]{ulem}
\usepackage[T1]{fontenc} % if needed
\usepackage{caption}

\newcommand{\be}{\begin{equation}}
\newcommand{\ee}{\end{equation}}
\def\bea{\begin{eqnarray}}
\def\eea{\end{eqnarray}}

\newcommand{\PP}{\mathcal{P}}
\newcommand{\T}{\mathcal{T}}
\newcommand{\calS}{\mathcal S}
\newcommand{\CcalS}{{\mathcal C}_{\mathcal S}}

\newcommand{\bfomega}{\mbox{\boldmath$\omega$}}

\title{\boldmath On systematic and GR effects on muon $g-2$ experiments}

\author[a]{Alessio Notari,}
\author[b,c]{Daniele Bertacca}

\affiliation[a]{Departament de F\'isica Qu\`antica i Astrofis\'ica \& Institut de Ci\`encies del Cosmos (ICCUB), Universitat de Barcelona, Mart\'i i Franqu\`es 1, 08028 Barcelona, Spain}
\affiliation[b]{Dipartimento di Fisica ed Astronomia, Universit\`a di Padova, Via Marzolo 8, 35131 Padova, Italy}
\affiliation[c]{INFN, Sezione di Padova, Via Marzolo 8, 35131 Padova, Italy}

\emailAdd{notari@fqa.ub.edu}
\emailAdd{daniele.bertacca@pd.infn.it}

\abstract{We derive in full generality the equations that govern the time dependence of the energy ${\mathcal E}$ of the decay electrons in a muon $g-2$ experiment. We include both electromagnetic and gravitational effects and we estimate possible systematics on the measurements of $g-2\equiv 2(1+a)$, whose experimental uncertainty will soon reach $\Delta a/a\approx 10^{-7}$. In addition to the standard modulation of ${\mathcal E}$ when the motion is orthogonal to a constant magnetic field $B$,  with angular frequency $\omega_a=e a |B|/m$, we study effects due to:  (1) a non constant muon $\gamma$ factor, in presence of electric fields $E$, (2) a correction due to a component of the muon velocity along $B$ (the ``pitch correction''), (3) corrections to the precession rate due to $E$ fields, (4) non-trivial spacetime metrics. Oscillations along the radial and vertical directions of the muon  lead to oscillations in ${\mathcal E}$ with a relative size of order $10^{-6}$, for the BNL $g-2$ experiment. We then find a subleading effect in the ``pitch'' correction, leading to a frequency shift of $\Delta \omega_a/\omega_a \approx {\cal O}(10^{-9})$ and subleading effects of about $\Delta \omega_a/\omega_a \approx {\rm few} \times {\cal O}(10^{-8}-10^{-9})$ due to $E$ fields. Finally we show that GR effects are dominated by the Coriolis force, due to the Earth rotation with angular frequency $\omega_T$,  leading to a correction of about $\Delta \omega_a/\omega_a \approx \omega_T/(\gamma \omega_a) \approx {\cal O}(10^{-12})$. A similar correction might be more appreciable for future electron $g-2$ experiments, being of order $\Delta \omega_a/\omega_{a, {\rm el}} \approx \omega_T/(\omega_{a, {\rm el}}) \approx 7\times 10^{-13}$, compared to the present  experimental uncertainty, $\Delta a_{\rm el}/a_{\rm el}\approx 10^{-10}$, and forecasted to reach soon  $\Delta a_{\rm el}/a_{\rm el}\approx 10^{-11}$.}

\begin{document}
\maketitle
\flushbottom

\section{Introduction}

A spinning particle orbiting in a plane orthogonal to a constant magnetic field experiences a precession of the spin, relative to its velocity, due to a gyromagnetic factor $g\equiv 2 (1+a) \neq 2$. The most recent measurements of $a$ for the muon~\cite{Bennett:2006fi} give $a_{EXP}= 116 592 091(63) \cdot 10^{-11} (3)$, and so have a relative precision of about $\Delta a/a\approx  5\times 10^{-7}$, while ongoing experiments should improve it by a factor of 3~\cite{Grange:2015fou}.
% and such a high precision is important because a discrepancy between theory and experiment might be a detection channel of new physics beyond the standard model.

The Standard Model prediction $a_{SM}$ has been computed~\cite{Keshavarzi:2018mgv, Davier:2017zfy, Jegerlehner:2017lbd} with an uncertainty
of about $3 \times 10^{-7}$. Any discrepancy from the experimental measurement could reveal additional contributions  beyond the Standard
Model~\cite{Miller:2012opa, Roberts:2010zz}. Using the hadronic contributions from~\cite{Keshavarzi:2018mgv, Davier:2017zfy, Jegerlehner:2017lbd}, there is indeed indication of a discrepancy 
between experiment and the Standard Model prediction, with a statistical significance
around $3.5-4$ standard deviations.

Some authors~\cite{Morishima:2018sae} have claimed non-negligible effects due to general relativity (GR) 
on the frequency of the muon spin precession in a magnetic field. This has been questioned by several other authors~\cite{Visser:2018omi, Guzowski:2018bjs, Nikolic:2018dap, Laszlo:2018llb, Venhoek:2018biz}.
In~\cite{Miller:2018jum} it has been shown that, even if the size of the GR corrections claimed in \cite{Morishima:2018sae} were correct, they would imply anyway a negligible effect, of order $\Delta a/a\approx 10^{-10}$,  for the Brookhaven National Laboratory (BNL) E821 experiment \cite{Bennett:2006fi}.
%
%The Standard Model prediction $a_{SM}$ has been theoretically calculated with an uncertainty
%of about ±0.3 ppm [10, 11, 12]. Should the experimental value [14],
%$a_{EXP}= 116 592 091(63) ? 10^?11 (3)$
%differ from the Standard Model value at a statistically significant level, it would reflect
%additional contributions from as yet undiscovered particles beyond those of the Standard
%Model [15, 16]. Using the hadronic contributions from Refs. [10, 11, 12], the difference
%between experiment and the Standard Model theory is positive, with a statistical significance
%between 3.5 and 4 standard deviations.

In the present paper we derive a full treatment of the spin precession in GR and we show what is the correct size of such effects in a realistic muon experiment. In order to do this, we need to clarify basic issues: what are the observable quantities and in which frame. What is actually observed in an experiment is the energy ${\cal E}$ of the decay products (electrons or positrons) of the muons, as seen in the laboratory frame. This will be shown to be given by the sum of two terms: one is proportional to the muon gamma factor and the second is the scalar product of a generalized spin $\Sigma$ with the observed velocity $v$.  In the particular case of flat spacetime and of a circular motion of muons of mass $m$, in a plane orthogonal to a constant magnetic $B$ field, the latter corresponds to the standard anomalous oscillatory term, with frequency $\omega_a = ea |B|/m$ ({\it e.g.} see \cite{Roberts-Lee-Marciano}).

It is well known that, even in the flat spacetime case, in a realistic experiment the muon is not on a perfect circular orbit but performs small radial and vertical oscillations. This implies: (1) a correction to the precession frequency, usually called ``pitch correction'' ({\it e.g.} see \cite{Farley:1973sc, Field:1974pe}); (2) corrections, due to electric fields, that are assumed to vanish at a particular value of the muon gamma factor (the {\it magic} momentum); (3) the fact that the muon gamma factor is not constant due to electric fields, which introduces additional time dependence of the electron energy ${\cal E}$.
We will analyze thus such corrections in the case of the BNL experiment. 

Then, we will analyze the extra terms introduced by the presence of a non-trivial metric, including the particular cases of a rotating metric and of a Schwarzschild metric, representing respectively Earth's rotation and gravity.

The paper is organized as follows: in section~\ref{sec:General} we introduce the general formalism, in section~\ref{sec:EOM} we write in full generality the equations of motion for velocity and spin. In section~\ref{sec:Setup} we define the quantities that are relevant for experiments and study the Minkowski cases with ({\it i}) pure magnetic field, ({\it ii}) magnetic plus electric fields. 
In section~\ref{sec:GR} we study then the effects of non-trivial metrics.
In section~\ref{sec:Experiments} we analyze the impact on a typical experiment of the various corrections and finally we draw our Conclusions. 

\label{sec:intro}

%
%	
%	\title{Coriolis effect on $g-2$ measurements}
%	
%
%%	\email{rferreira@icc.ub.edu}
%	\author{Alessio Notari$^{1}$}
%	\email{notari@fqa.ub.edu}
%	
%	\affiliation{$^{1}$ Departament de F\'isica Qu\`antica i Astrofis\'ica \& Institut de Ci\`encies del Cosmos (ICCUB), Universitat de Barcelona, Mart\'i i Franqu\`es 1, 08028 Barcelona, Spain}
%	
%	\author{Daniele Bertacca}
%	\affiliation{Dipartimento di Fisica e Astronomia ``G. Galilei'', Universita' degli Studi di Padova, via Marzolo 8, I-35131,
%Padova, Italy}
%\affiliation{INFN, Sezione di Padova, via Marzolo 8, I-35131, Padova, Italy}
%\affiliation{INAF-Osservatorio Astronomico di Padova, Vicolo dell'Osservatorio 5, I-35122 Padova, Italy}
%
%	\begin{abstract}
%		
%
%
%	\end{abstract}
%	
%	
%	\maketitle
	
	\section{General setup} \label{sec:General}
	
	We use the spacetime signature $(-,+,+,+)$ and we will work with a generic metric $g_{\mu\nu}$ (unless specified), so that scalar products between any two 4-vectors, $A$ and $B$, are given by
	$A\cdot B=A^\mu g_{\mu \nu} B^\nu$. We mostly follow the treatment and notation of~\cite{Felice:2010cra}.
	
	A given frame (F) is defined by the observer 4-velocity $u^{\rm (F)}$. In the rest of this paper we will use  the muon rest frame, ${\rm (F)=(M)}$, and the laboratory frame, ${\rm (F)=(L)}$, although we will drop the $L$ index in the latter case.
	The energy per unit mass of a particle with 4-velocity $U$, measured by $u^{\rm (F)}$, is its gamma factor, given by
	\begin{eqnarray}
	\gamma^{\rm (F)}=-(U\cdot u^{\rm (F)}) \label{gamma} \, .
	\end{eqnarray}
	Four-velocities are always normalized, $U\cdot U= u^{\rm (F)}\cdot u^{\rm (F)}=-1$. 
	
	The time observed in the frame $(F)$ is given by
   \begin{eqnarray}
        dT^{\rm (F)}=\gamma^{\rm (F)} d\tau \, ,
        \end{eqnarray}
	where $\tau$ is the proper time of the particle with velocity $U$.
	
	The observer in a given (F) frame measures a velocity $v^{\rm (F)}$ for the above particle, given by:
	\begin{eqnarray}
	v^{\rm (F)}=\frac{U}{\gamma^{\rm (F)}}-u^{\rm (F)} \label{v} \label{vUeq} \, .
	\end{eqnarray}
	By construction $v^{\rm (F)}$ is purely spatial with respect to $u^{\rm (F)}$, {\it i.e.}  $v^{\rm (F)}\cdot u^{\rm (F)}=0$, and $|v^{\rm (F)}|=\sqrt{v^{\rm (F)}\cdot v^{\rm (F)}}$ is the modulus of the observed velocity. Indeed the following property holds:
	\begin{eqnarray}
	\gamma^{\rm (F)}=1/\sqrt{1-|v^{\rm (F)}|^2} \,	. 
	\end{eqnarray}
 	
In general, useful quantities are the projector operators, which define the decomposition of the local frame of a given observer $u$ into orthogonal subspaces. Precisely we define the temporal and spatial projector operators in the following way:
\begin{eqnarray} 
{\T(u)}=-u\otimes u \quad \quad {\rm and} \quad \quad  {\PP(u)}=I + u\otimes u \, ,
\end{eqnarray}
or in terms of their components:
\begin{eqnarray} 
{\T(u)}^\mu_\nu=-u^\mu u_\nu \quad \quad {\rm and} \quad \quad {\PP(u)}^\mu_\nu=\delta^\mu_\nu + u^\mu u_\nu\; \, .
\end{eqnarray}

% 	let us define the projector $\PP(u)$  on a vector $A$, as
%		\begin{eqnarray} 
%	(\PP(u) A)^\mu \equiv \PP^\mu_\nu A^\nu= (\delta^\mu_\nu+u^\mu u_\nu)  A^\nu \, .
%		\end{eqnarray}

	The spin of a particle is described by a 4-vector $S$, with the property that it is orthogonal to its 4-velocity:
	\begin{eqnarray}
	S\cdot U=0 \, .
	\end{eqnarray}

It is also useful to define the spatially projected spin, $\Sigma^{\rm (F)}=\PP(u^{\rm (F)})  S $, on a given observer. This satisfies then the following equations
	\begin{eqnarray}
S=\Sigma^{\rm (F)}- (S \cdot u^{\rm (F)}) u^{\rm (F)} = \Sigma^{\rm (F)}+(\Sigma^{\rm (F)}\cdot v^{\rm (F)}) u^{\rm (F)} \label{eqsigma} \, ,
\end{eqnarray}
where we have used that
\begin{eqnarray}
\Sigma^{\rm (F)}\cdot v^{\rm (F)} = - S \cdot u^{\rm (F)} \, . 
\end{eqnarray}
Note that indeed $\Sigma^{\rm (F)}$ is spatially projected  with respect to $u^{\rm (F)}$, since $\Sigma^{\rm (F)}\cdot u^{\rm (F)}=0$.

	In the particle rest frame (M) the observer velocity is $U$ itself and so we simply have 
	\begin{eqnarray}
	\Sigma^{\rm (M)}=S + (S\cdot U) U=S  \,. \label{spinrest}
	\end{eqnarray}
	In this  frame we may use, for example, coordinates $\{x_{\rm(M)}^{\hat \mu}\}$ defined in the adapted frame $\{e_{\rm (M)}^{\hat \mu}\}$, where $U=e^{\rm (M)}_{\hat 0}$, $e^{\rm (M)}_{\hat \alpha } \cdot e^{\rm (M)}_{\hat \beta} = \eta_{\hat \alpha  \hat \beta}$ and $e^{\rm (M)}_{\hat 0} \cdot e_{\rm (M)}^{\hat a}=0$, where $\eta_{\hat \alpha  \hat \beta}\equiv {\rm diag}(-1,1,1,1)$. Then we find\footnote{$S^{\hat{\mu}}$ is briefly analyzed in curved space in Appendix \ref{Sec:calS}.} %,  o we have $U^{\hat{\mu}}=(u^{\hat{0}},0,0,0)$, so the spin takes the form
	\begin{eqnarray}
	\Sigma^{\hat \mu}_{\rm (M)}=S^{\hat \mu}_{(\rm (M)}=e_{)\rm M)}^{\hat \mu} \cdot S\equiv (0,s^{\hat{m}}) \, , \qquad {\rm with} \, \qquad |S|^2=S\cdot S = {\bf s}\cdot{\bf s} \, . \label{spinrest}
	\end{eqnarray}
We use Latin indices, from 1 to 3, and boldface for three-dimensional vectors and we defined ${\bf s}\cdot{\bf s}=s^{\hat m} s^{\hat n} \eta_{\hat m \hat n}$, where $\eta_{\hat m  \hat n}\equiv e^{(\rm M)}_{\hat m} \cdot e^{(\rm M)}_{\hat n}$. Greek indices are always from 0 to 4.

	\section{Equations of motion} \label{sec:EOM}

Let us now consider a particle (the muon) in an electromagnetic field, described by an antisymmetric field strength $F$, and in a generic metric. We need the equation of motion for the observed velocity and spin. 

We first review the definitions of electric and magnetic fields as observed by the laboratory frame, with 4-velocity $u^{(L)}\equiv u$.  In any  system of coordinates the  fields observed by $u$ are described by 4-vectors $E$ and $B$,
 \begin{eqnarray}
	\, E^\alpha= F^{\alpha \rho} u_\rho \,  \qquad  B^\alpha=\frac{1}{2} \eta^{\rho \alpha \mu \nu} F_{\mu \nu} u_\rho \, , \qquad {\rm with} \qquad 
	\eta^{\rho \alpha \mu \nu} \equiv \frac{1}{\sqrt{-g}} \epsilon^{\rho  \alpha \mu \nu} \, ,
	\end{eqnarray}
where $g={\rm det}(g_{\mu \nu})$ and $ \epsilon_{\rho  \alpha \mu \nu}$ is the standard totally antisymmetric Levi-Civita symbol, $ \epsilon_{0 1 2 3}= -\epsilon^{0 1 2 3} =1$.
Note that both $E$ and $B$ are purely spatial w.r.t. $u$, that is $E\cdot u=B\cdot u=0$. 
One can also write the $F$ tensor as:
   \begin{eqnarray}
   F_{\beta\mu}=u_{\beta} E_\mu-u_{\mu} E_\beta+\eta(u)_{\beta \mu}^{\,\,\,\,\,\,\,\sigma} B_\sigma \, , \label{effe}
   \end{eqnarray}
 	  where 
	  \[\eta(u)_{\alpha \beta \gamma}\equiv u^\delta \eta_{\delta \alpha \beta \gamma} \quad  {\rm with} \quad \eta_{\delta \alpha \beta \gamma} = \sqrt{-g}\, \epsilon_{\delta \alpha \beta \gamma}\;.\]% with $g=Det(g_{\mu \nu})$.
	     
Let us check the flat space case, with an observer at rest in Minkowski coordinates: $u^\mu=(1,0,0,0)$. In our signature we have $ u_\mu=(-1,0,0,0) \, ,$ and so:
   \begin{eqnarray}
	 E^1 &=& - F^{1 0} \, ,\qquad  B^1 = \frac{1}{2} \epsilon^{\rho 1 \mu \nu} F_{\mu \nu} u_\rho=\frac{1}{2} \epsilon^{1 m n} F_{m n} =  F_{23}\, , \nonumber \\
	E^2 &=& - F^{2 0} \, ,\qquad B^2  =   F_{31} \, ,\nonumber\\
	E^3 &=& - F^{3 0} \, ,\qquad B^3  =   F_{12} \, .
	\end{eqnarray}
Using the above definition of the Levi-Civita symbol, here we have used $\epsilon^{mnp}\equiv -\epsilon^{0  mnp}$ and  $ \epsilon_{mnp}\equiv \epsilon_{0  mnp}$.

\subsection{Velocity} 
We will consider now also the motion of muons as seen by the  laboratory frame (L), defined by a 4-velocity $u$. A muon with 4-velocity $U$ has therefore in this frame an observed velocity $v \equiv v^{\rm (L)}$ and a gamma factor $\gamma\equiv \gamma^{\rm (L)}$.

The equation of motion for the muon velocity is
	\begin{eqnarray}
	\frac{D U}{D\tau}=\frac{e}{m} F\ast U \label{eom} \, ,
	\end{eqnarray}
	where our notation is $(F\ast U)^\mu \equiv F^{ \mu  \nu} U_{ \nu }$ in a given coordinate system, $e$ is the electric charge and $m$ is the muon mass.
	We will use the covariant derivative of a four-vector $A$ always along $U$, defined as 	
	\begin{eqnarray} \label{k}
	\frac{D A}{D\tau}=\left(\frac{d A}{d\tau}\right)^\alpha + [\Gamma A U]^\alpha \, ,
	\end{eqnarray}
	where we used the shorthand notation $[\Gamma A B]^\mu=\Gamma^{\mu}_{ \nu \rho } A^{ \nu} B^{\rho}$, with $\Gamma^{\mu}_{ \nu \rho } $ the Christoffel symbols in a given coordinate system and $A$ and $B$ any two vectors. We can also decompose the equation of motion using the observed $E$ and $B$ fields\footnote{See~\cite{Felice:2010cra}, Eqs.~(6.133)-(6.136).}:
	\[\frac{D U}{D\tau} =\T(u) \frac{D U}{D\tau} +\PP(u) \frac{D U}{D\tau}\;,\]
	We have thus
	\begin{equation} 
	\PP(u) \frac{D U}{D\tau} =\gamma f_{\rm EM} \, ,
	\end{equation}
  where $f_{\rm EM}$ is just the Lorentz force\footnote{See Eq.~(6.75) of Ref.\,\cite{Felice:2010cra}.}:
	\begin{equation}
	f_{\rm EM}= \frac{e}{m} \left[E+(v \times_u B) \right] \label{gammadot} \, .
	\end{equation}
Here we have introduced the spatial cross product  $\times_u$ in the following way
	\begin{equation} [A\times_u B]^\alpha \equiv \eta(u)^{\alpha}_{\beta \gamma} A^\beta B^\gamma\;. \label{cross}
		\end{equation}
	So we find
\begin{eqnarray} 
\left(\frac{D U}{d\tau}\right)^\alpha = \frac{e}{m} \gamma \left[E^\alpha+(v \times_{u} B)^\alpha \right] - \frac{e}{m} u^\alpha (u\ast F\ast U) \, ,\label{Lorentz1}
\end{eqnarray}
where  $u\ast F\ast U\equiv u_\mu F^{\mu \nu} U_\nu$.

%\textcolor{red}{Define $\times_u$} 
	Now let us write an equation for $dv/dT$, {\it i.e.} the evolution of the observed velocity w.r.t. the observed time $T$. Using Eq.~(\ref{vUeq}) we have that:
	\begin{eqnarray}
	\frac{dv}{dT}= \frac{1}{\gamma^2} \frac{d U}{d\tau} - \frac{U}{\gamma^2} \frac{d\gamma}{dT} -\frac{du}{dT} \, .
	\label{veq0}
	\end{eqnarray}
Now we can use the following relation\footnote{See Eq.~(6.81)  of Ref.\,\cite{Felice:2010cra}.}, 
	\begin{eqnarray}
	\dot{\gamma}\equiv d\gamma/dT=v\cdot (f_{\rm EM}+f_{\rm G}) = v\cdot \left( \frac{e}{m} E+f_{\rm G}\right) \, , \label{gammadot}
	 	\end{eqnarray}
	where he have defined the gravitational force as \cite{Felice:2010cra} :
	\begin{eqnarray} f^\alpha_{\rm G} \equiv - P(u)^\alpha_\beta \left(\frac{D u}{D \tau}\right)^\beta  = -  \left(\frac{D u}{D \tau}\right)^\alpha = -   \left(\frac{d u}{d \tau}\right)^\alpha -[\Gamma u U]^\alpha \,. \label{FG}
		\end{eqnarray}
%		where we have introduced used the definition of the projector $\PP (u)$ on a vector $A$.
%		, as
%		\begin{eqnarray} 
%	(P(u) A)^\mu \equiv P^\mu_\nu A^\nu= (\delta^\mu_\nu+u^\mu u_\nu)  A^\nu \, .
%		\end{eqnarray}
%	The second term is zero  because $u$ is unitary, so we just have:
%		\begin{eqnarray} F^{G\, , \mu} = - \frac{D u^\mu}{D \tau} = -  \frac{d u}{d \tau} -[\Gamma u U]
%			\end{eqnarray}

\subsection{Spin}
Let us write now the equation for the spin \cite{Bargmann:1959gz} of a particle with gyromagnetic factor $g$, adapted to our signature ($-,+,+,+$), as:
   \begin{eqnarray}\label{EqSpin}
	\frac{D S}{d\tau} &=& \frac{g e}{2m} \left[ F\ast S - (S \ast F \ast U)  U \right] + \left( \frac{D U}{ d\tau} \cdot S \right) U = \label{BMT1} \nonumber \\
	&=&  \frac{g e}{2m} ( F\ast S) - \frac{e}{m}\ a  (S \ast F \ast U)  U \, ,
	\end{eqnarray}
	where $a\equiv \left( \frac{g}{2} - 1 \right) $.
	%(*The last term of the first line (with a plus sign) coincides with Bini, Eq.~10.1 and Weinberg. The terms proportional to $g$ are instead absent in Bini.*)
	One can check that signs are consistent by evaluating the flat space case with both observer and particle at rest, $u^\mu=U^\mu=(1,0,0,0)$, and with\footnote{In such a case we have that 
	 \begin{eqnarray}
	\frac{d {\bf s}^a}{d\tau} &=&  \frac{g e}{2m} ({\bf s}\times {\bf B})^a \, ,
	%  \\
%	\frac{d S^0}{d\tau} &=& \left( \frac{d U}{ d\tau} \cdot S \right) \, .
		\end{eqnarray}
	where boldface are used for three dimensional vectors in Minkowski, ${\bf B}$ is the spatial part of $B$ and $\times$ is the standard three-dimensional vector product in flat space. We will also use in the following $\cdot$ for the standard three-dimensional vector product in flat space.} $S=(0,{\bf s})$. 	
	
	Let us now work in the most general case and let us write  an equation for $\Sigma^{\rm(L)} \equiv \Sigma$. Eq.~(\ref{eqsigma}) becomes
\begin{eqnarray}
S=\Sigma+(\Sigma\cdot v) u=\Sigma- (S \cdot u) u \, .  \label{SSigma}
\end{eqnarray}
and inserting this in the l.h.s. of Eq.~(\ref{BMT1}) we get\footnote{See Ref.~\cite{Felice:2010cra}, Eq.~(10.8), or Ref.~\cite{Jantzen:1992rg}.}:
	 \begin{eqnarray}
	\frac{D S}{d\tau} = \frac{D {\Sigma}}{d\tau} +  \frac{D u}{d\tau} (v\cdot {\Sigma}) + u \frac{D}{D\tau}(v\cdot \Sigma) \, .
	\end{eqnarray}
	Now we can spatially project this w.r.t. $u$, 
%	and get:
%	 \begin{eqnarray}
%	P(u)\frac{D {\Sigma}}{d\tau} +  P(u)\frac{D u}{d\tau} (v\cdot {\Sigma}) =   \frac{g e}{2m} P(u) (F\cdot S) -  \frac{e}{m} a \, P(u) U (S \cdot F \cdot U) 
%	\end{eqnarray}
% Now use the definition of $P(u)$
together with  Eq.~(\ref{FG}), so that\footnote{This coincides with the first line of Eq.~(10.15) by \cite{Felice:2010cra}, plus an additional term proportional to $g$.}:
 \begin{eqnarray} 
	P(u)\frac{D {\Sigma}}{d\tau} 
	%\equiv \frac{D_{({\rm fw}, U, u)} {\Sigma}}{d\tau}
	= f_G (v\cdot {\Sigma})   +  \frac{g e}{2m} P(u) (F\ast S) -   \frac{e}{m} a \gamma v (S \ast F \ast U) \, .\label{eqsigma2}
		\end{eqnarray}
%	where we also used the definition Eq.~(3.157) by Bini book (since ${\Sigma}$ is orthogonal to $u$).

We may also write it in terms of the observed time $T$,
 \begin{eqnarray}
	 P(u)\left( \gamma \frac{d {\Sigma}}{dT} + [\Gamma \Sigma U] \right) = f_G (v\cdot {\Sigma})   + \frac{g e}{2m} P(u) (F\ast S) -   \frac{e}{m} a \gamma v (S \ast F \ast U) \,  .
	\end{eqnarray}
	We can now use Eq.~(\ref{effe}) and get:
	 \begin{eqnarray}
	  P(u) (F\ast S)^\alpha &=& P^{\alpha \beta} F_{\beta \mu} S^\mu= P^{\alpha \beta}(u_\beta E_\mu-u_\mu E_\beta+\eta_{\beta \mu}^{\,\,\,\,\,\sigma} B_\sigma) ({\Sigma}+(v\cdot {\Sigma}) u)^\mu = \nonumber \\
	  & = & E^\alpha(v\cdot {\Sigma})+ P^{\alpha \beta}\eta_{\beta \mu}^{\,\,\,\,\,\sigma} B_\sigma {\Sigma}^\mu \;,+P^{\alpha \beta} \eta_{\beta \mu}^{\,\,\,\,\,\sigma} B_\sigma u^\mu (v\cdot {\Sigma}) =
	  \nonumber \\
	  & = & E^\alpha(v\cdot {\Sigma}) +  ({\Sigma}\times_u B)^\alpha\;, \label{FS}
	\end{eqnarray}
 where we used
%	  By definition we have:
	  	 \begin{eqnarray}
		  E_\beta u^\beta=0\;, \quad  \eta_{\beta \mu}^{\,\,\,\,\,\sigma} B_\sigma {\Sigma}^\mu = ({\Sigma}\times_u B)_\beta \, , \quad	u^\beta \eta_{\beta \mu}^{\,\,\,\,\,\sigma} B_\sigma {\Sigma}^\mu &=&0  \quad  {\rm and } \quad \eta_{\beta \mu}^{\,\,\,\,\,\sigma} B_\sigma u^\mu =0 \, . \nonumber
	  	 \end{eqnarray}
%So we get:
% \begin{eqnarray}
%	  P(u) (F\cdot S)^\alpha &=&  E^\alpha(v\cdot {\Sigma})+ ({\Sigma}\times_u B)^\alpha  \, .	
%	  \label{FS} \end{eqnarray}	
	Similarly:
	 \begin{eqnarray}
	(S \ast F\ast U) &=&  S^\beta F_{\beta \mu} U^\mu=  ({\Sigma}^\beta+(v\cdot {\Sigma}) u^\beta) (u_\beta E_\mu-u_\mu E_\beta+\eta_{\beta \mu}^{\,\,\,\,\,\sigma}  B_\sigma) \gamma (u^\mu+v^\mu) = \nonumber \\
%	&=& \gamma \left[ ({\Sigma}\cdot E) - (v\cdot {\Sigma}) (E \cdot v) \right] +  \gamma (v\cdot {\Sigma}) u^\beta \eta_{\beta \mu}^\sigma B_\sigma v^\mu +  {\Sigma}^\beta \eta_{\beta \mu}^\sigma B_\sigma U^\mu \\
	&=& \gamma \left[ ({\Sigma}\cdot E) - (v\cdot {\Sigma}) (E \cdot v) \right] +  \gamma {\Sigma} \cdot (v \times_u B) \, . \label{SFU1}
	\end{eqnarray}

\section{Experimental setup} \label{sec:Setup}

In an experimental setup muons are moving on an (almost) circular orbit and the laboratory is at rest with the earth surface. Muons decay then into a neutrino-antineutrino pair and an electron (we use the word electron to represent either the positron or electron in the generic $\mu \rightarrow e\nu \nu$ decay chain) and the observable quantity is the number of electrons whose energy in the (L) frame is above a given threshold. For simplicity we will look only at electrons emitted with maximal energy  $\mathcal{E}_{{\rm Max}}$ in the (M) frame: 
\begin{eqnarray}
   \mathcal{E}_{{\rm Max}}^{\rm (M)}  =  \frac{m^2+m_e^2-\left(m_{\bar \nu_e}+m_{\nu_\mu}\right)^2}{2m}\;.
\end{eqnarray}
This happens when a pair of neutrino-antineutrino is emitted in the same direction and the electron is thus emitted in the opposite direction. In this case the neutrino-antineutrino pair carries zero angular momentum and so the electrons inherits the same spin of the muons. In the limit of $m_e \rightarrow 0$ only one helicity can be produced and thus electron emission is zero (or maximal) if its velocity is aligned (or anti-aligned) with its spin, which is equal to the muon spin. Therefore the emission of electrons with maximal energy is peaked when the velocity of the electron is opposite to the muon spin direction. %(or, actually, opposite to it).

We can idealize and simplify this situation  by assuming that:
\begin{itemize}
\item In the muon rest frame (M), where the observer velocity is the muon 4-velocity $U$, all electrons are emitted with the same energy $\mathcal{E}^{\rm (M)}=m_e \gamma_e^{\rm (M)}$.  %and  so the same modulus of the momentum 
%$|p_e^{\rm (M)}|=m_e \gamma_e^{\rm (M)} |v_e^{\rm (M)}|$.
Here $\gamma_e^{\rm (M)}$ and $v_e^{\rm (M)}$ are the electron gamma factor and velocity, as observed by $U$. Using Eqs.~(\ref{gamma}) and (\ref{v}) they read:
\begin{eqnarray}
	\gamma_e^{\rm (M)}=-(U_e\cdot U) \, ,\\
	v_e^{\rm (M)}= \frac{U_e}{\gamma_e^{\rm (M)}}-U \, ,
	\end{eqnarray}
	where $U_e$ is the electron 4-velocity.
	
\item We assume that the electrons in the frame (M) are emitted with velocity $v_e^{\rm (M)}$ opposite to the muon spin direction $\Sigma^{\rm (M)}$:
\begin{eqnarray}
\hat{v}_e^{\rm (M)}=-\hat{\Sigma}^{\rm (M)} \, ,
\end{eqnarray}
where
\begin{eqnarray}
\hat{v}_e^{\rm (M)} \equiv \frac{v_e^{\rm (M)}}{|v_e^{\rm (M)}|} \, , \, \qquad \hat{\Sigma}^{\rm (M)} \equiv \frac{\Sigma^{\rm (M)}}{|\Sigma^{\rm (M)}|} =\frac{S}{|S|} \, .
\end{eqnarray}
\end{itemize}
We will always use a hat for vectors normalized to 1.
Now we can express $U_e$ in the two frames as:
\begin{eqnarray}
U_{e}&=&\gamma_e^{\rm (L)} (u+v_e^{\rm (L)}) = \gamma_e^{\rm (M)} (U+v_e^{\rm (M)}) \, ,  \qquad  {\rm where} \qquad \gamma_e^{\rm (L)}=-(U_e\cdot u) \, . 
%\\ U_{e}&=&\gamma_e^{\rm (M)} (U+v_e^{\rm (M)})  \, ,  \qquad   {\rm where} \qquad  \gamma_e^{\rm (M)}=-(U_e\cdot U) \, .
\end{eqnarray}
So let us now compute the experimentally observed quantity, %\footnote{Here$\mathcal{E} = \mathcal{E}^{\rm (M)}$}
 $\mathcal{E}/m_e=\gamma_e^{\rm (L)}$, using the above equations and also Eq.~(\ref{SSigma}):
\begin{eqnarray}
\gamma_e^{\rm (L)} &=& - (U_e\cdot u) = -\gamma_e^{\rm (M)} (U+v_e^{\rm (M)})\cdot u =\
%&=& -\gamma_e^{\rm (M)} (-\gamma + v_e^{\rm (M)} \cdot u) = \\
%&=& \gamma_e^{\rm (M)} \left( \gamma +  \frac{|v_e^{\rm (M)}|}{|S|} (S \cdot u) \right) = \\
=  \gamma_e^{\rm (M)} \left[ \gamma -  \frac{|v_e^{\rm (M)}|}{|S|} (\Sigma \cdot v) \right] \, . \label{energyel}
\end{eqnarray}
Here all quantities are constant in time, except for $\gamma$ (the muon gamma factor) and $(\Sigma \cdot v)$ (proportional to the alignment between the muon spin and the muon velocity in the (L) frame).
In the following we will write the above equation as
\begin{eqnarray}
\mathcal{E}=\mathcal{E}_0 + \mathcal{E}_R \, , \label{omegaeq}
\end{eqnarray} 
where
\begin{eqnarray}
\mathcal{E}_0 \equiv   m_e \gamma_e^{\rm (M)} \gamma , \qquad \mathcal{E}_R \equiv  - \mathcal{E}_1 (\Sigma \cdot v) =   \mathcal{E}_1 (S \cdot u) \, , \qquad
\mathcal{E}_1\equiv  m_e \gamma_e^{\rm (M)} \frac{|v_e^{\rm (M)}|}{|S|}  \, . \label{omegazero}
\end{eqnarray}
We will now write an equation for $\dot{\mathcal{E}}$ in various setups: ({\it i}) in Minkowski metric, with the observer at rest and with a static magnetic field only, ({\it ii})  in Minkowski metric, with the observer at rest and with both electric and magnetic fields, ({\it iii}) including a non-trivial metric and motion of the observer.
\subsection{Static magnetic field in flat space}
Let us first work in a simplified framework, where we assume: 
\begin{itemize}
\item $B$ constant in time in the (L) frame; 
\item No electric fields in the (L) frame ($E = 0$);
\item Minkowski metric and observer at rest $u^\mu=(1,0,0,0)$.
\end{itemize}
The $\gamma$ of the muons is thus constant in this case, from Eq.~(\ref{gammadot})
%The electron energy can be written as
%\begin{eqnarray}
%\omega_e^{\rm (L)}=\mathcal{E}_0 + \omega(T)=\mathcal{E}_0 + \mathcal{E}_1 (\Sigma \cdot v) =\mathcal{E}_0 - \mathcal{E}_1 (S \cdot u)  \, , \label{energyeM}
%\end{eqnarray}
and so $\mathcal{E}_0$ and $\mathcal{E}_1$ in Eq.~(\ref{omegazero}) are constants. 
In this simplified framework the time-dependence of the observed electron energy is only due to $\mathcal{E}_R$. As we will see,  under the condition $v\cdot B=0$, this turns out to have a periodic behavior with respect to the observer's time $T$, with a frequency $\omega_a$, the anomalous precession frequency.
%One possibility is to compute $\omega_a$ in the following way:
%\begin{eqnarray}
%\omega^2_a = - \frac{1}{E(t)}  \frac{d^2 \mathcal{E}}{dT^2} 
%\end{eqnarray}
Let us indeed compute time derivatives of the electron energy as follows:
\begin{eqnarray}
\frac{1}{\mathcal{E}_1}  \frac{d \mathcal{E}}{dT} =\frac{d S}{dT}  \cdot  u =    \frac{d S}{\gamma d\tau} \cdot  u \, .
\end{eqnarray}
Now we use Eq.~(\ref{BMT1}) and  get:
\begin{eqnarray}
 \frac{d S}{d\tau} \cdot u &=&  \frac{g e}{2m}  u \cdot ( F\ast S)  - \frac{e}{m} a (S \ast F \ast U)   U \cdot  u =  \frac{e}{m} a \gamma (S \ast F \ast U) \, \, ,
\end{eqnarray} 
since the first term vanishes in absence of electric fields.
Using Eq.~(\ref{SFU1}) and keeping again only the magnetic part we find:
%\begin{eqnarray}
% u\cdot \frac{d S}{d\tau} &=& \frac{e}{m} \alpha \gamma^2  {\Sigma} \cdot (v \times_u B)
%\end{eqnarray} 
%leading thus to:
\begin{eqnarray}
\frac{1}{\mathcal{E}_1} \frac{d \mathcal{E}}{dT} = - \frac{e}{m} a \gamma   B \cdot (v \times_u {\Sigma}) \, .
\label{dEdT}
\end{eqnarray}
 From here we immediately see that there is a non-trivial time dependence only if $g\neq2$.
 
 %We may also write the above equation as
%\begin{eqnarray}
%\frac{d (\Sigma \cdot v)}{dT} &=&  \frac{e}{m} a \gamma   B \cdot (v \times \Sigma) \, , \label{omegavec} 
%\end{eqnarray}
%One can also define the precession frequency of $v$ and $\Sigma$ as
%\begin{eqnarray}
%\frac{d \hat{v}}{dT} &=&  \vec{\omega}_c \times \hat{v} \, , \label{omegavec} \\
%\frac{d \hat{\Sigma}}{dT} &=&  \vec{\omega}_\Sigma \times \hat{\Sigma} \
%\end{eqnarray}
%which imply that
%\begin{eqnarray}
%\frac{d (\hat{s}\cdot \hat{v})}{dT} &=&  (\vec{\omega}_v- \vec{\omega}_s) \cdot (\hat{v}\times \hat{s}) \, , \label{omegavec} 
%\end{eqnarray}

%\textcolor{red}{This part may not hold in GR, check in a generic metric}
Now, the spin $S$ can be written in the rest frame coordinates as in Eq.~(\ref{spinrest}), or it can be written also using the (L) coordinates ({\it e.g.} see~\cite{Jackson:1998nia}, chapter 11), as:
\begin{eqnarray}
S= \left(\gamma {\bf v}\cdot {\bf s},{\bf s}+\frac{\gamma^2}{\gamma+1} ({\bf v}\cdot {\bf s}) {\bf v} \right) 
\label{Slab} \, ,
\end{eqnarray}
where ${\bf v}$ is the spatial part of $v$ in the L coordinates. One can check explicitly that in both coordinates $S\cdot U=0$ and $S^2 = S\cdot S= {\bf s}\cdot {\bf s}$.

This implies that the scalar and vector products are equal to
\begin{eqnarray}
\Sigma \cdot v &=&-S\cdot u= \gamma \, {\bf s}\cdot {\bf v} \label{Sigmas1}   \, , \\
\Sigma \times_u v&=&  {\bf s} \times {\bf v} \label{Sigmas2} \, ,
\end{eqnarray}
so that using
\begin{eqnarray}\label{E_e^L}
\frac{\mathcal{E}-{\mathcal E}_0}{\mathcal{E}_1}=-\gamma \, {\bf s}\cdot {\bf v} \, ,
\end{eqnarray}
together with Eq.~(\ref{dEdT}) we have
\begin{eqnarray}
 \frac{d ( {\bf s}\cdot {\bf v})}{dT} =  \frac{e}{m} a    [{\bf B} \cdot ({\bf v} \times {\bf s})] \, . \label{spinanomaly}
\end{eqnarray}
where ${\bf B}$ is the spatial part of $B$. This coincides with the magnetic term of Eq.~(9) by Bargmann et al.~\cite{Bargmann:1959gz} and Eq.~(11.171) of~\cite{Jackson:1998nia}. Note, though, that such a form of the equations mixes quantities in different frames.

It is also customary to define the following quantities:
\begin{eqnarray}
\frac{d \hat{\bf v}}{dT} &\equiv &  \bfomega_{\bf v} \times \hat{\bf v} \, , \label{omegavecV} \\
\frac{d \hat{\bf s}}{dT} &\equiv &  \bfomega_{\bf s}  \times \hat{\bf s} \   \label{omegavecS} \, .
\end{eqnarray}
The first quantity can be found combining Eqs.~(\ref{Lorentz1}), (\ref{veq0}), and the second combining Eqs.~(\ref{SSigma})-(\ref{SFU1}) and~(\ref{Slab}), leading to\footnote{Here we have also used the following identity 
\[A\times_u [B\times_u (C\times_u  D)]=B[A\cdot_u (C\times_u  D)]-(A \cdot_u  B)(C\times_u  D)\;. \]
We can also find Eqs. (\ref{omegavecVEXPL}) and (\ref{omegavecSEXPL})  in~\cite{Jackson:1998nia}.}:

%\DB{Dobbiamo espandere con piu' dettagli la derivazione di queste equazioni}
\begin{eqnarray}
 \bfomega_{\bf v} &=&- \frac{e}{m \gamma} {\bf B} \, , \label{omegavecVEXPL} \\
 \bfomega_{\bf s} &=&- \frac{e}{m \gamma} {\bf B} - \frac{e}{m} a \left[ {\bf B} -\frac{\gamma}{\gamma+1} ({\bf v} \cdot {\bf B}) {\bf v} \right] \, , \label{omegavecSEXPL}
\end{eqnarray}
in agreement with~\cite{Jackson:1998nia}.
The above equations also imply that
\begin{eqnarray}
\frac{d (\hat{\bf s}\cdot \hat{\bf v})}{dT} &=&  ( \bfomega_{\bf v} - \bfomega_{\bf s}) \cdot (\hat{\bf v}\times \hat{\bf s}) \, , \label{omegavec} 
\end{eqnarray}
One can indeed define a quantity 
\begin{eqnarray}
 \tilde{\bfomega}_a &\equiv& \bfomega_{\bf v} - \bfomega_{\bf s} = \frac{e}{m} a \left[ {\bf B} - \frac{\gamma}{\gamma+1}  ({\bf v} \cdot {\bf B}) {\bf v}  \right] \, , \label{omegawithpitch}
\end{eqnarray}
where the second term is the so-called ``pitch correction'',
\begin{eqnarray}
\bfomega_{\rm P} \equiv - \frac{e}{m} a  \frac{\gamma}{\gamma+1}  ({\bf v} \cdot {\bf B}) {\bf v}  \, . \label{pitch}
\end{eqnarray}
One should be aware, though, that the quantity $\tilde{\bfomega}_a$ does not have a clean physical interpretation, since it is a difference of two vectors defined in different frames. Indeed we will see  in subsection~\ref{general} that $| \tilde{\bfomega}_a|$ does {\it not} represent the oscillation frequency in a fully correct way.

Note also that only the component of $\tilde{\bfomega}_a$ orthogonal to the plane defined by $\hat{\bf s}$ and $\hat{\bf v}$ enter in Eq.~\ref{omegavec}, so that it reduces to
\begin{eqnarray}
\frac{d (\hat{\bf s}\cdot \hat{\bf v})}{dT} &=& \bfomega_a \cdot (\hat{\bf v}\times \hat{\bf s}) \, , \label{omegavec2} 
\end{eqnarray}
where
\begin{eqnarray}
 \bfomega_a  \equiv \frac{e}{m} a {\bf B} \, . \label{omegaadef}
\end{eqnarray}

\subsubsection{Solving the equations of motion}

If the constant magnetic field is homogenous in space, one can write it as ${\bf B}=(0,0,B_\perp)$.  The velocity rotates thus only in the $x$-$y$ plane, so it has the form ${\bf v}=(v_x(t),v_y(t),v_z)$, where $v_z={\rm const}$, because of Eq.~(\ref{omegavecV}) and Eq.~(\ref{omegavecVEXPL}) . 

Let us study first the case of a particle with an initial velocity orthogonal to the magnetic field, $v\cdot B=0$. In this case $v_z=0$ at all times and so the spin ${\bf s}$ also rotates  only in the $x$-$y$ plane, as
${\bf s}=(s_x(t),s_y(t),s_z)$, with $s_z= {\rm const.}$, because of Eq.~(\ref{omegavecS}) and Eq.~(\ref{omegavecSEXPL}).

From Eq.~(\ref{spinanomaly}) one can show that in this case the angular frequency of oscillation of $\hat{\bf v}\cdot \hat{\bf s}$ is $\omega_a\equiv |\bfomega_a|$. Indeed one can define the following decomposition:
\begin{eqnarray}
\hat{\bf s}=|s_{||}| (\cos(\omega_a T) \hat{\bf v} + \sin(\omega_a T) \hat{\bf n} ) + s_z \hat{\bf z}\, , \label{anomalous}
\end{eqnarray}
where $\hat{\bf n}$ is a unit vector orthogonal to $\hat{\bf v}$ lying in the $x$-$y$ plane and $s_{||}=$ constant, with $S^2=s_z^2+s^2_{||}$. 
As a consequence we will have an oscillatory behavior of $\hat{\bf s}\cdot\hat{\bf v}$ and of $\mathcal{E}(T)$:
\begin{eqnarray}
\frac{\mathcal{E}(T)}{\gamma |v| |S| \mathcal{E}_1} &=& \hat{\bf s}\cdot \hat{\bf v} = \cos(\omega_a T)  \, , \label{anomalous} \\
\frac{\dot{\mathcal{E}}(T)}{\gamma |v| |S| \mathcal{E}_1}&=& -\omega_a \sin(\omega_a T) \, . \label{eqBcirc}
\end{eqnarray}
%\textcolor{red}{where:
%\begin{eqnarray}
%\omega_a =  \frac{e}{m} a  |B| \label{anomalous} \, ,
%\end{eqnarray}
%with $B_\perp\equiv {\bf B}\cdot (\hat{\bf n}\times \hat{\bf v})$.}

If instead $v_z=v\cdot B \neq 0$ then the  evolution of $s$ is more complicated, due to the last term of Eq.~(\ref{omegavecSEXPL}),  the pitch correction, and the motion is not simply harmonic.

One can also infer the frequency from the following definition:
\begin{eqnarray}
r_{v_z=0,\,{\bf E}={\bf 0}}\equiv \frac{1}{\mathcal{E}_1 \gamma [\hat{B} \cdot (\Sigma \times_u v)]}\frac{d \mathcal{E}}{dT} =\frac{1}{\mathcal{E}_1 \gamma [\hat{\bf B} \cdot ({\bf s}\times {\bf v})]}\frac{d \mathcal{E}}{dT} =  \frac{e}{m} a  B_\perp = \omega_a \, .
\end{eqnarray}

Note that Bargmann  {\it et al.}~\cite{Bargmann:1959gz}  defined an angle $\phi$ as follows:
\begin{eqnarray}
&& S/|S| = \cos(\phi) e_l +  \sin(\phi) e_t \, , \\
&& e_t=(0,\hat{\bf n})\, , \qquad
e_l=\gamma(|v|,\hat{\bf v})   \, , \\
&& e_t\cdot e_t = 1 \, , \qquad e_l\cdot e_l=1 \, , \qquad e_l \cdot e_t=0 \, , 
\end{eqnarray}
where $L$ coordinates have been used to define $e_l$ and $e_t$.
Using Eqs.~(\ref{Slab}), (\ref{Sigmas1}) and (\ref{Sigmas2}) we have that
\begin{eqnarray}
|S| \sin(\phi)=S\cdot e_t= {\bf s}\cdot \hat{\bf n}= \frac{[\hat{B} \cdot (\Sigma \times_u v)]}{|v|}  \, ,\\
|S| \cos(\phi)=S\cdot e_l= {\bf s}\cdot \hat{\bf v}=  \frac{(\Sigma\cdot v)}{|v| \gamma} \, .
\end{eqnarray}
They used, then,  its time derivative as a definition of the anomalous precession frequency:
%\begin{eqnarray}
%\dot{\phi}&=&-\frac{1}{\sin{\phi}} \frac{d\cos\phi}{dT} = -\frac{1}{{\bf s}\cdot \hat{\bf n} } \frac{d}{dT} ( {\bf s}\cdot \hat{\bf v})= -\frac{1}{{\bf s}\cdot \hat{\bf n} } \frac{d}{dT} \left(\frac{\Sigma\cdot v}{v \gamma} \right)= \\
%&=& \frac{1}{\gamma |\Sigma \times_u v|} \left[   \frac{d}{dT} (\Sigma\cdot v) +  (\Sigma\cdot v) \gamma v   \frac{d}{dT} \left( \frac{1}{v \gamma}\right)\right] = \\
%%&=&  \frac{1}{\gamma |\Sigma \times_u v|}  \frac{d}{dT} (\Sigma\cdot v)  +   \frac{(\Sigma\cdot v) }{\gamma |\Sigma \times_u v|}  \frac{\gamma}{1-\gamma^2}\dot{\gamma} \\
%&=& r +   \frac{(\Sigma\cdot v) }{\gamma |\Sigma \times_u v|}  \frac{\gamma}{1-\gamma^2} \dot{\gamma} \, .
%\label{dotphi}
%\end{eqnarray}
\begin{eqnarray}
\dot{\phi}&=&-\frac{1}{\sin{\phi}} \frac{d\cos\phi}{dT} = -\frac{1}{{\bf s}\cdot \hat{\bf n} } \frac{d}{dT} ( {\bf s}\cdot \hat{\bf v}) \, .
\label{dotphi}
\end{eqnarray}

Note that, crucially the quantity $\dot{\phi}$ coincides with our definition $r$ only when $\gamma$ is constant. As we will see in the next subsection $\dot{\phi}$ and $r$ do {\it not} coincide in  the presence of electric fields aligned with the velocity since $\dot{\gamma}=(e/m) v\cdot E$.%, as we will see in the next subsection.

\subsubsection{General treatment} \label{general}

The oscillation frequency $\omega$ can be found in a more general way, by computing the second derivative  of $\mathcal{E}$, starting from Eq.~(\ref{dEdT}):
\begin{eqnarray}
-\omega^2 \equiv \frac{1}{\mathcal{E}}\frac{d^2 \mathcal{E}}{dT^2} =  \frac{e}{m (\Sigma\cdot v)} a \gamma   B \cdot \left( \frac{d v}{d T} \times_u {\Sigma}+ v \times_u \frac{d {\Sigma}}{dT} \right) \, .
\end{eqnarray}
Using Eq.~(\ref{Lorentz1}) and Eq.~(\ref{veq0}) we get:
\begin{eqnarray}
 \frac{d v}{d T} \times_u {\Sigma} = \frac{e}{m \gamma} (v \times_u B) \times_u {\Sigma} =  \frac{e}{m \gamma} (B({\Sigma}\cdot v)- v({\Sigma}\cdot B)) \, ,
\end{eqnarray}
and using Eq.~(\ref{BMT1}), together with Eq.~(\ref{SSigma}) and with the property that $v\cdot u=v\times_u u=0$, we get:
\begin{eqnarray}
 v\times_u \frac{d {\Sigma} }{d T}  = \frac{g e}{2 m \gamma} v\times_u ({\Sigma} \times_u B) =  \frac{g e}{2 m \gamma} [{\Sigma} (v \cdot B) - B({\Sigma}\cdot v)] \, ,
\end{eqnarray}
arriving thus at:
%\begin{eqnarray}
% \frac{d^2 \mathcal{E}}{dT^2} = - \mathcal{E}_1 \frac{e^2}{m^2} a^2  B^2  ({\Sigma}\cdot v) +  \mathcal{E}_1  \frac{e^2}{m^2}   a^2  (B\cdot {\Sigma}) (v \cdot B) \, .
%\end{eqnarray}
%So the anomalous frequency is given by:
\begin{eqnarray}
 \omega^2 =  \frac{e^2}{m^2} a^2  B^2  \left[ 1 -   \frac{(\hat{B}\cdot \hat{{\Sigma}}) (\hat{v} \cdot \hat{B})}{\hat{\Sigma} \cdot \hat{v}} \right] \, . \label{secondderivative}
\end{eqnarray}
Note that this result has been computed in a completely coordinate-free way. 

Now, the above quantity $\omega$ is not simply a number, but a function of time $T$.
If $B\cdot v=0$ it reduces to a number, meaning that we have an exact harmonic motion with a constant frequency, in agreement with Eq.~(\ref{anomalous}) and (\ref{eqBcirc}).
In the case $B\cdot v\neq 0$ we get instead a correction. Note that this differs from the so-called ``pitch correction'' of Eq.~(\ref{omegawithpitch}). 
Indeed using Eq.~(\ref{Slab}) the above equation becomes:
\begin{eqnarray}
 \omega^2 =  \frac{e^2}{m^2} a^2  B^2  \left\{ 1 - \left[ \frac{\gamma}{\gamma+1} ({\bf  v}\cdot {\bf \hat B})^2+\frac{({\bf \hat B}\cdot {\bf \hat s})({\bf \hat v} \cdot {\bf \hat B})}{\gamma ({\bf \hat v}\cdot {\bf \hat s})} \right] \right\} \, , \label{finalomega2}
\end{eqnarray}
while  from Eq.~(\ref{omegawithpitch}) we have that
\begin{eqnarray}
|\tilde{\bfomega}_a|^2=\tilde{\bfomega}_a\cdot\tilde{\bfomega}_a =  \frac{e^2}{m^2} a^2  B^2  \left[ 1 - (v\cdot \hat{B})^2 \right] \, . \label{omegaPsquared} 
\end{eqnarray}
At large $\gamma$ the two expressions are very similar to each other, but we will check  in section~\ref{sec:Experiments} that Eq.~(\ref{finalomega2}) better reproduces numerical results.

\subsection{Electric and magnetic field  in flat space}

In this section we still keep Minkowski metric and observer at rest, but we reintroduce the electric fields.  In this case, when taking the derivative of Eq.~(\ref{omegaeq}), $\gamma$ is not constant, from Eq.~(\ref{gammadot}), leading to
\begin{eqnarray}
 \frac{1}{\mathcal{E}_1}\frac{d \mathcal{E}}{dT}\Big|_{e.m.}   =  \frac{e}{m} \frac{|S|}{|v_e^{\rm (M)}|}  (E \cdot v) + u \cdot \frac{d S}{\gamma d \tau} \, .
\end{eqnarray}
Now using Eq.~(\ref{BMT1}) we  obtain the full electromagnetic contributions:
\begin{eqnarray}
 \frac{1}{\mathcal{E}_1}\frac{d \mathcal{E}}{dT}\Big|_{e.m.}
&=&  \frac{e}{m} \frac{|S|}{|v_e^{\rm (M)}|} (E \cdot v) -  \frac{e}{m \gamma} \left( \frac{g}{2} u\cdot(F\ast S) + a \gamma (S\ast F\ast U) \right) = \label{eqE} \nonumber \\
&=& \frac{e}{m} \frac{|S|}{|v_e^{\rm (M)}|}  (E \cdot v) + \frac{e}{m \gamma} \left(   \frac{g}{2}  {\Sigma}\cdot E - a \gamma^2 \left[ ({\Sigma}\cdot E) - (v\cdot {\Sigma}) (E \cdot v)  +   {\Sigma} \cdot (v \times_u B) \right] \right) = \nonumber \\
&=& \frac{e}{m} \Bigg[ \frac{|S|}{|v_e^{\rm (M)}|}  (E \cdot v) -   a \gamma   (v\cdot {\Sigma}) (E \cdot v)   +    a \gamma   B \cdot ( {\Sigma} \times_u v) -   \frac{(1-a (\gamma^2-1))}{\gamma} ({\Sigma}\cdot E) \Bigg] \, .\nonumber \\  \label{fullE}
\end{eqnarray}
Now, a general definition of the parameter $r$, which can be used also in this situation, is the following
\begin{equation}\label{r}
r \equiv  \frac{1}{\mathcal{E}_1 \gamma |v| (S \cdot e_t)}\frac{d \mathcal{E}|_{e.m.}}{dT} \;.
\end{equation}
This leads to
\begin{eqnarray}\label{r-rel}
r  &=& \frac{e}{m} a \frac{ B \cdot ( {\Sigma} \times_u v)}{|v| (S\cdot e_t)}   + \frac{e}{m \gamma} \frac{|S|}{|v_e^{\rm (M)}|}  \frac{(E \cdot v)}{|v| (S\cdot e_t)} - \frac{e}{m} a  \frac{ (v \cdot {\Sigma}) (E \cdot v) }{|v| (S\cdot e_t) } -   \frac{e}{m \gamma^2} \frac{{\Sigma}\cdot E }{|v| (S\cdot e_t)} [1-a (\gamma^2-1)]  \, .\nonumber\\
 \label{finalMINK}
\end{eqnarray}
The first term is the one we analyzed in the previous subsection.  The last term vanishes at the so-called {\it magic} momentum $\gamma=(1+a^{-1})^{1/2}$. 
The second and third  terms are the ones due to $\dot{\gamma}$. 
We will analyze the impact of such terms in experiments in section~\ref{sec:Experiments}.
%\textcolor{red}{In particular the third term relative to the first  has roughly a size $\gamma ~ E\cdot v/B_\perp$ }. The second term has a relative size 
%$1/(a \gamma) (E\cdot \hat{v})/B_\perp \approx  \gamma  (E\cdot \hat{v})/B_\perp $ at the {\it magic} momentum, assuming ultrarelativistic electrons and muons.

Note that if we used instead $\dot{\phi}$, as defined in Eq.~(\ref{dotphi}), we would get:
\begin{eqnarray} \label{phidot}
\dot{\phi}%&=& r +   \frac{(\Sigma\cdot v) }{\gamma |\Sigma \times_u v|}  \frac{\gamma}{1-\gamma^2} \dot{\gamma} =  r +   \frac{e}{m} \frac{(\Sigma\cdot v) (E\cdot v)}{|\Sigma \times_u v|}  \frac{1}{1-\gamma^2} = \\
&=&  \frac{e}{m} a \frac{ B \cdot ( {\Sigma} \times_u v)}{|v| (S\cdot e_t)}  + \frac{e}{m}   \frac{ (v\cdot {\Sigma}) (E \cdot v) }{|v| (S\cdot e_t)} \left(a- \frac{1}{\gamma^2-1} \right) - \frac{e}{m \gamma^2} \frac{{\Sigma}\cdot E }{|v| (S\cdot e_t)} [1-a (\gamma^2-1)] \, , \nonumber\\
\end{eqnarray}
so that also the second and the third term would vanish at the magic momentum. However the quantity directly observable in an experiment is the period of variation of the energy, which is set by $r$, not $\dot{\phi}$. Therefore the choice of the magic momentum does not guarantee that effects due to electric fields can be completely neglected.

Note also that the first additive term on the {l.h.s} of Eq.~(\ref{phidot}) can be written  in a different way, {\it  i.e.}, taking into account that
\[\frac{ B \cdot ( {\Sigma} \times_u v)}{|v| (S\cdot e_t)} = \frac{|S| \sin(\phi) B \cdot ( e_t \times_u v)}{|v| (S\cdot e_t)}= {\bf v}\cdot ({\bf B} \times \hat{\bf n})\;, \]
where we used that ${\Sigma} \times_u v=S \times_u v$ and $S\cdot e_t = |S| \sin(\phi)$.  In this case, using the above relation,  Eq.~(\ref{phidot}) exactly coincides with the result obtained in Bargmann  {\it et al.} \cite{Bargmann:1959gz}.
%The fourth term instead comes just from the time dependence of $\omega_0$ in Eq.~\ref{omegaeq}.

Finally, also in this case, one may compute the second derivative of the energy ${\cal E}$, leading to a lengthy result:
\begin{eqnarray}
-\omega^2\equiv \frac{1}{{\cal E}_1} \frac{d^2 {\cal E}  }{dT^2} &=& \frac{e^2}{m^2} 
\Bigg\{ \frac{|S|}{|v_e^{(M)}|}  \left( \frac{E^2+B\cdot(E\times v)-(E\cdot v)^2}{\gamma}+\dot{E}\cdot v  \right)  
+ \nonumber \\
 &+&  a^2 B^2 (\Sigma\cdot v) -a^2 (B\cdot v)  (B\cdot \Sigma) \nonumber \\
 &-& (E\cdot v)^2 (\Sigma\cdot v) a (1+a)  + a^2 B\cdot(E\times v) (\Sigma\cdot v) \nonumber \\
 &+&   a (1+a) B\cdot(\Sigma \times v) (E\cdot v)   +\frac{B\cdot(E\times \Sigma) (-1 -  2 a +a^2 (\gamma^2-1) )  }{ \gamma^2}  \nonumber \\
 &+& \frac{E^2 (\Sigma\cdot v) (-1 -  2 a +a^2 (\gamma^2-1) )  }{ \gamma^2} + \frac{(E\cdot \Sigma) (E\cdot v) (1 + a +a \gamma^2) }{ \gamma^2}   \nonumber \\
  &-& a \gamma (\Sigma\cdot v) (\dot{E}\cdot v) -  \frac{1-a(\gamma^2-1)}{\gamma}  (\Sigma \cdot \dot{E}) +a \gamma \dot{B}\cdot(\Sigma\times v) \Bigg\} \, . \label{lengthy}
\end{eqnarray}
Here the time derivatives of the $E$ and $B$ field have to be computed along the trajectory as
\begin{eqnarray}
\dot{E}^i=\partial_j E^i v^j \, , \qquad \dot{B}^i=\partial_j B^i v^j \, .
\end{eqnarray}
In the BNL experiment~\cite{Bennett:2006fi} there is an electric field (of quadrupolar form) leading to a non-trivial to $\dot{E}$.
If the magnetic field can be considered perfectly homogeneous then $\dot{B}=0$. As we have stressed, not all the terms vanish at the magic momentum.
We will comment on the impact of such terms in section~\ref{sec:Experiments}.

\section{Effects due to a generic GR metric} \label{sec:GR}

The full result for the derivative of the electron energy, in presence of a generic motion of $u$ in a generic metric is:
\begin{eqnarray}
\frac{1}{\mathcal{E}_1}\frac{d \mathcal{E}}{dT} &=&  \frac{m_e \gamma_e^{\rm (M)}}{\mathcal{E}_1} \dot{\gamma} +   \left( u \cdot \frac{D S}{d T} + \frac{D u}{dT}\cdot S \right) = \nonumber \\ 
&=&    \frac{|S|}{|v_e^{\rm (M)}|}  \left(\frac{e}{m}  E +f_G \right) \cdot v +  u \cdot \frac{D S}{d T}  -  \frac{1}{\gamma}  \left(f_G \cdot S \right) = \\
&=& \frac{1}{\mathcal{E}_1}\frac{d \mathcal{E}}{dT}\Big|_{e.m.} +\frac{|S|}{|v_e^{\rm (M)}|}  f_G \cdot v -  \frac{1}{\gamma}  \left(f_G \cdot S \right) \, .
\label{genericmetric}
\end{eqnarray}
where we used Eq.~(\ref{gammadot}) and~(\ref{FG}).
The first term $(1/\mathcal{E}_1) d \mathcal{E}/dT|_{e.m.} $ is exactly as in Eq.~(\ref{fullE}), where all the scalar products have to be taken with the full metric and all the vector products as $\times_u$, defined in Eq.~(\ref{cross}). Then, there are purely gravitational terms: one arising from $\dot{\gamma}$, proportional to $ f_G \cdot v$, and a new term proportional to $f_G\cdot S$.

We will study the impact of such terms in two specific cases: a rotating observer and a Schwarzschild metric.

Before doing that, we will also write separate equations for $\hat{\Sigma}$ and for $\hat{v}$ in the following subsections.

%
%In the rotating case we may adopt coordinates of the lab,  $u=(u^0,0,0,0)$. Moreover we can work at linear order in the Earth's angular frequency $\omega$, since this is a tiny effect.
%The expression of the Christoffel symbols and $u^0$ of this case is given in Appendix. 
%Both  $ \frac{d u}{d\tau}$ and $[\Gamma u U]$ are quadratic in $\omega$, so we may neglect them.
%So the only non-trivial quantity is 
%\begin{eqnarray}
%%f_G \cdot v &=& -r v^r \gamma \gamma_T^2 \omega^2 \\
% \frac{1}{\gamma}  u \cdot [\Gamma S U] &=& \Omega \cdot (v \times_u S) \, ,
%\end{eqnarray}
%where $\Omega\equiv (0,0,0,\omega)$.
%Comparing with Eq.~\ref{omegavec}, the precession frequency gets corrected at linear order in $\omega$  as follows:
%\begin{eqnarray}
% \bfomega_a =  \frac{e}{m} a {\bf B}-\frac{\bfomega}{\gamma} \, .
%\end{eqnarray}

\subsection{Equations for $\hat{\Sigma}$ and $\hat v$}

We are going to write here the time derivative of the electron energy, in presence of a generic motion of $u$ in a generic metric, in terms of differential equations for $\hat{\Sigma}$ and $\hat{v}$.
Here we define $\hat{\Sigma}$ and $\hat{v}$ in the following way:
\be \label{hatsigmav}
\hat{\Sigma} = {\Sigma \over |\Sigma|} \quad \quad {\rm and} \quad \quad \hat{v}={v \over |v| }\,,
\ee
where
\be
|\Sigma| =  \left[\Sigma \cdot \Sigma \right]^{1/2}  \quad \quad {\rm and} \quad \quad   |v| = \left[v \cdot  v \right]^{1/2}  \;.
\ee
Starting again with the energy of the electron in the laboratory frame
\be
\mathcal{E}=m_e  \gamma_e^{\rm (M)} =   m_e \gamma_e^{\rm (M)} \left[\gamma - \frac{|v_e^{\rm (M)}|}{|S|}  |\Sigma|\,|v| \left(\hat\Sigma \cdot \hat v\right)\right]\;.% \mathcal{E}_0 + \mathcal{E}_R \, , \label{omegaeq} \, ,
\ee
Defining the spatial momentum per unit mass of the electron as $p \equiv \gamma v\, $, we have that
\be
{d | p | \over d T}=(f_{\rm EM}+f_G)\cdot v = \gamma {d | v | \over d T} + \dot \gamma | v |=\gamma {d | v | \over d T}  +  v\cdot \left( f_{\rm EM}+f_{\rm G}\right) | v |\; ,
\ee
and so
\be
 {d | v | \over d T} = \frac{\left(f_{\rm EM}+f_G\right)\cdot \hat v}{\gamma^3}.
\ee
Multiplying Eq. (\ref{eqsigma2}) by $\hat{\Sigma}$ and using Eq. (\ref{hatsigmav},a), after a little algebra we can also compute the time derivative of $|\Sigma|$, %\DB{Dobbiamo espandere con piu' dettagli la derivazione di queste equazione}
%\be
% {d | \Sigma | \over d T} = \left(\Sigma \cdot v\right) \mathcal{F} \;,
%\ee
%where 
%\be
%\mathcal{F}=\frac{(1+a)e}{m_e\gamma}(E\cdot\hat\Sigma)+{1\over\gamma}(f_G\cdot\hat\Sigma)- a\gamma \left[(f_{\rm EM} \cdot\hat\Sigma) - (v\cdot\hat\Sigma) (f_{\rm EM}\cdot v)\right]\;.
%\ee
\be \label{|Sigma|}
 {d | \Sigma | \over d T} = \left(\Sigma \cdot v\right) \left[\mathcal{F}\cdot \hat \Sigma +  {e a\gamma \over m}  |v| \,  B \cdot \left(\hat v \times_u \hat \Sigma \right) \right]\;,
\ee
where the four vector $\mathcal{F}$ is defined as
\be
\mathcal{F}=\frac{(1+a)e}{m\gamma}E+{1\over\gamma}f_G- {e a\gamma \over m} \left[E  -  (E \cdot v) v \right]\,.
\ee
%and
%\be
%\tilde \Omega_{\mathcal{E}} =  B\;.
%\ee
We can also rewrite Eq.~(\ref{|Sigma|}) in different way. Indeed, taking into account that\footnote{We have used the following identity
\[\left(A \times_u B \right)\left(C \times_u D \right)=(A \cdot_u C) (B \cdot_u D) - (A \cdot_u D) (B \cdot_u  C)\;,\]
where $\cdot_u= P(u)$.
}
\be 
\left(\Sigma \cdot v\right) (\mathcal{F}\cdot \hat \Sigma) =  | v | | \Sigma | \left[\left(\hat \Sigma \times_u \mathcal{F} \right) \cdot \left(\hat v \times_u \hat \Sigma \right) + \mathcal{F}\cdot \hat v  \right]\;,
\ee
 Eq.~(\ref{|Sigma|}) becomes
 \be \label{|Sigma|2}
 {d | \Sigma | \over d T} =  | v | | \Sigma |\left\{\left[\left(\hat \Sigma \times_u \mathcal{F} \right) +   {e a\gamma \over m}  |v| \left(\hat\Sigma \cdot \hat v \right)  \,  B   \right] \cdot \left(\hat v \times_u \hat \Sigma \right) + \mathcal{F}\cdot \hat v    \right\}\;.
\ee
Another important ingredient is
\be
 {d | \hat v \cdot  \hat \Sigma | \over d T} =  \Omega_{ a} \cdot \left(\hat v \times_u \hat \Sigma\right)\;,
\ee
with $ \Omega_{ a}\equiv \Omega_v - \Omega_\Sigma$, where  $\Omega_v $ are $ \Omega_\Sigma$ are  respectively the angular velocity of $v$ and $\Sigma$, which will be defined and computed explicitly in the next subsections. Adding all the above relations we find

\bea \label{dE/dT}
{d \mathcal{E} \over d T}&=& \gamma_e^{\rm (M)}\left[v\cdot \left( f_{\rm EM}+f_{\rm G}\right) \right] -  \gamma_e^{\rm (M)}\frac{|v_e^{\rm (M)}|}{|S|} \Bigg\{{| v |}^2 | \Sigma | \left(\hat v \cdot \hat \Sigma\right) (\mathcal{F}\cdot \hat v) 
 \nonumber\\ 
 &&+ {| \Sigma | \over \gamma^3}\left[\hat v\cdot \left( f_{\rm EM}+f_{\rm G}\right) \right] \left(\hat v \cdot \hat \Sigma\right) + | v | | \Sigma | \left[ \Omega_{\rm tot} \cdot \left(\hat v \times_u \hat \Sigma\right)\right]
\Bigg\}\,,
\eea
where
\be
\Omega_{\rm tot} \equiv  \Omega_{ a} + \Omega_{\mathcal{E}} = \Omega_v - \Omega_\Sigma + \Omega_{\mathcal{E}}\;,
\ee
with
\be
\Omega_{\mathcal{E}}
 \equiv  | v | \left(\hat\Sigma \cdot \hat v \right) \left[\left(\hat \Sigma \times_u \mathcal{F} \right) +   {e a\gamma \over m}  |v| \left(\hat\Sigma \cdot \hat v \right)  \,  B   \right]\;.
 %{e a\gamma \over m_e} {| v |}^2 \left(\hat v \cdot \hat \Sigma\right)^2 \, B\;.
\ee
In order to obtain $\Omega_{\rm tot}$,  let us note that if we consider a generic four-vector $X$ orthogonal to $u$, {\it i.e.} $X\cdot u=0$, we can write 
\be 
P(u) {D X \over d T}= {D_{({\rm fw},U,u)}X\over dT} = \left({D_{({\rm fw},U,u)} X \over dT} \right)^{\hat \mu} e_{\hat \mu}\;,
\ee
 where, still for   $X\cdot u=0$, we have defined  the ``Fermi-Walker total spatial covariant derivative'' in the following way
\be
 \gamma {D_{({\rm fw},U,u)}X\over dT} =   {D_{({\rm fw},U,u)}X\over d\tau} \equiv P(u)  {D X\over d\tau}\;.% = P(u)P(U) \nabla_U X\;.
\ee
%Here the Fermi-Walker derivarive $ {D_{({\rm fw},U)}X / d\tau} $ on a generic four-vector $Y$ is defined in the following way
% \[
% {D_{({\rm fw},U)}Y\over d\tau} =  \nabla_U Y + a(U) (U \cdot Y) - U (a(U) \cdot Y).
% \]

Now, we  are considering the frame  $\{e_{\hat \mu}\}$ which is the orthonormal frame adapted to the observer $u$, so that $X\cdot u=X \cdot e_{\hat 0}=X_{\hat 0}=0$. Applying ${D_{({\rm fw},U,u)} / dT} $ to $\{e_{\hat a}\}$, we have\footnote{From the definition of $\{e_{\hat a}\}$, we immediately note that 
\[
  {D_{({\rm fw},u)}e_{\hat a} \over dT} = 0\;.\]
  This is the reason why this tetrad frame is also called Fermi-Walker.
  Here we have defined the Fermi-Walker derivative $ {D_{({\rm fw}, Z)}Y / d\tau_Z} $ on a generic four-vector $Y$ along the time-like vector $Z$ ({\it i.e.} $Z \cdot Z=-1$) in the following way
 \[
 {D_{({\rm fw},Z)}Y\over d\tau_Z} \equiv {D Y \over d\tau_Z} + a(Z) (Z \cdot Y) - Z (a(Z) \cdot Y)\,,
 \]
where 
\[a(Z)\equiv {D Z \over d\tau_Z}\;.\]
Obviously, in our case, if $Z=\{U,u\}$ then $\tau_Z=\{\tau,T\}$\;.} \cite{Jantzen:1992rg, Felice:2010cra}
 \be \label{e-fw}
 {D_{({\rm fw},U,u)} e_{\hat a}\over dT} = \left[\zeta_{({\rm fw})}+\zeta_{({\rm sc})}\right] \times_u  e_{\hat a}\;,
 \ee
where
$\zeta_{({\rm fw})}$ is the so-called Fermi-Walker angolar velocity vector
\be
\zeta_{({\rm fw)}}^{\hat a}= -{1\over 2}\eta(u)^{\hat a \hat b \hat c} C_{({\rm fw})\hat b \hat c}\;,
\ee
$C_{({\rm fw})\hat a \hat b}$ are termed Fermi-Walker structure functions,
\[C_{({\rm fw})\hat b \hat a} \equiv e_{\hat b} \cdot \left[P(u)  {D e_{\hat a}\over d T} \right]\;,\]
and $\zeta_{({\rm sc})}$ is the spatial curvature angular velocity\footnote{
Here we have defined
\[\Gamma_{[\hat c | \hat d | \hat b]}\equiv {1 \over 2 } \left(\Gamma_{\hat c  \hat d  \hat b} - \Gamma_{\hat b  \hat d  \hat c} \right)\;.\]
}, 
\be
\zeta_{({\rm sc})}^{\hat a} \equiv  -{1\over 2}\eta(u)^{\hat a \hat b \hat c} \Gamma_{[\hat c | \hat d | \hat b]} v^{\hat  d }\;.
\ee

As we will see in the next subsections this adapted frame is useful in order to get quickly $\Omega_v$ and $\Omega_\Sigma$, which are measured in the observed frame (see also the discussion in Refs. \cite{Jantzen:1992rg, Felice:2010cra}).

\subsubsection{Analytical expression for $\Omega_v$}
Applying $D_{({\rm fw},U,u)} /d T$ on $p$ we have
\be
{D_{({\rm fw},U,u)}  p  \over d T} =(f_{\rm EM}+f_G)={d | p | \over d T} \hat v + {D_{({\rm fw},U,u)}  \hat v  \over d T}  | p |\;,
\ee
and using 
\be
{d | p | \over d T} = \left[(f_{\rm EM}+f_G) \cdot \hat v \right]\;,
\ee
we get
\be
{D_{({\rm fw},U,u)}  \hat v  \over d T}  = {(f_{\rm EM}+f_G) \over |p| }- {d \ln | p | \over d T} \;  \hat v = { (f_{\rm EM}+f_G) - \left[ (f_{\rm EM}+f_G) \cdot \hat v \right] \hat v \over  | p | } \, ,
\ee
or, equivalently,
\be
\left({d  \hat v  \over d T}\right)^{\hat a} = \left[\Omega_v \times_u \hat v\right]^{\hat a}\;,
\ee
where
\be
\Omega_v=-{\left(f_{\rm EM}+f_G\right) \times_u \hat v \over |p| } - (\zeta_{({\rm fw})}+\zeta_{({\rm sc})})\;.
\ee
Here we have used Eq.~(\ref{e-fw}).

\subsubsection{Analytical expression for $\Omega_\Sigma$}

Let us rewrite Eq.~(\ref{eqsigma2}) as
% differentiate along $U$ both sides of  Eq.~(\ref{eqsigma}), {\it i.e. }
%\be\label{Diff-eqsigma}
%{D S   \over d \tau}=  {D \Sigma   \over d \tau} -f_G (v \cdot \Sigma ) + u  {d    \over d \tau} (v \cdot \Sigma ) \, .
%\ee
%Then, using Eqs.~(\ref{eqsigma2}), (\ref{FS}), (\ref{SFU1}) and acting on both sides with $P(u)$, from Eq.~(\ref{Diff-eqsigma}) we find
\bea \label{Sigma-fw}
{D_{({\rm fw},U,u)}  \Sigma  \over d \tau}  &=& P(u)  {D \Sigma\over d\tau} = {(1+a)e\over m} \left[(v \cdot \Sigma) E + \Sigma \times_u B\right]+ (v \cdot \Sigma) f_G -a\gamma^2 (f_{\rm EM} \cdot \Sigma) v \nonumber\\
&&+ {ae \over m} \gamma^2 (v \cdot \Sigma)  (E \cdot v) v\;.
\eea
Now, using the definition of $\hat \Sigma $, this leads to
\bea \label{hatSigma-fw}
{D_{({\rm fw},U,u)} \hat \Sigma  \over d \tau}  &=& - \frac{d \ln |\Sigma|}{d \tau} \, \hat \Sigma +  {(1+a)e\over m} \left[(v \cdot \Sigma) E + \Sigma \times_u B\right]+ (v \cdot \Sigma) f_G -a\gamma^2 (f_{\rm EM} \cdot \Sigma) v \nonumber\\
&&+ a \gamma^2 (v \cdot \Sigma)  (f_{\rm EM} \cdot v) \,v\;.
\eea
From Eq.~(\ref{|Sigma|2}) and taking into account of Eq.~(\ref{e-fw}),  the above equation can be rewritten in the following way
\be 
\left({d  \hat \Sigma \over d T}\right)^{\hat a} = \left[\Omega_\Sigma \times_u \hat v\right]^{\hat a}\;,
\ee
where
\bea
\Omega_\Sigma &=& \hat \Sigma \times_u \Bigg\{ \frac{(1+a)e}{m \gamma}\left[(v \cdot \hat\Sigma)E+\left( \hat \Sigma  \times_u  B \right)\right]+ \frac{1}{\gamma}(v \cdot \hat\Sigma)f_G
\nonumber\\
&&+\gamma a \left[(f_{\rm EM} \cdot\hat\Sigma) - (v\cdot\hat\Sigma) (f_{\rm EM}\cdot v)\right]v\Bigg\} - (\zeta_{({\rm fw)}}+\zeta_{({\rm sc})})\;.
\eea

Then, in order to have a complete analysis of the problem, in  appendix \ref{Sec:calS}, we will generalize the three-vector ${\bf s}$ in a general GR frame.  

\subsection{Gravitational terms}

We study quantitatively here the size of the gravitational terms, focusing on two particular cases of interest: (i) Minkowski spacetime with a rotating observer, with angular frequency $\omega_T$; (ii) Schwarzschild metric, with observer at rest at some radial coordinate $r$.
This allows us to analyze separately the two effects of rotation and gravity of the Earth. Clearly the leading effects will be order ${\cal O}(\omega_T)$ and of order ${\cal O}(g)$ respectively, where $g\equiv M G/r^2$ is the gravitational acceleration felt at a fixed coordinate $r$, $M$ is the Earth's mass and $G$ is Newton's constant. We will disregard any cross terms between these two effects.

The only ingredients that we need are the metric inside the scalar and vector products and an expression for $f_G$, which can be readily evaluated using Eq.~(\ref{FG}), inserting the Christoffel symbols in a chosen coordinate system.

Finally we will also study, as a simple extension, the case of a gravitational wave metric.

\subsubsection{Rotating metric}

In this subsection we neglect Earth's gravity and consider only effects due to its rotation, at an angular frequency $\omega_T \simeq 7 \times 10^{-5} {\rm Hz}$. 
	The dominant effect will be a Coriolis acceleration, which is of order $|\omega_T\times v|\approx \omega_T$ (we work in units $c=1$). This has to be compared with $\omega_a\equiv a \cdot (e B/m) \approx  10^6   {\rm Hz} $, giving a relative error of order $\omega_T/\omega_a \approx 5\times 10^{-11}$. This is about 4 orders of magnitude smaller than the present experimental uncertainty for muon experiments, but we compute it anyway, since it represents the leading effect due to a nontrivial metric and it might be important for future experiments. Note also that factors of $\gamma$ here have to be computed, as they may change the above rough estimate, given that $\gamma\approx 29.3$.
	
	It is also clear that the Coriolis acceleration is much larger than the acceleration due to gravity on the surface of the Earth, since $|\omega_T\times  v|/g\approx 2\times 10^3$, which is the reason why Earth's gravity can be completely neglected.

	We start from an inertial reference system, described by coordinates $x_I^\mu=(t,x^i)$, whose metric is simply $$ds^2=-dt^2+dx_i dx^i=-dt^2+dr^2+r^2d\phi^2+dz^2, $$ where we employ also cylindrical coordinates.
	We consider then a system that rotates at an angular frequency $\omega_T$, {\it i.e.}, described by $\varphi\equiv \phi-\omega_T t={\rm const}$. The metric can thus be written as
	%\footnote{In this letter, the Greek indices take the values $0,1,2,$ and $3$ (referring to space-time coordinates) and the Latin indices take only the values $1,2$ and $3$ (referring to the spatial coordinates).}
\begin{eqnarray}
	ds^2=-dt^2 (1-\omega_T^2 r^2)&+&dr^2+r^2d\varphi^2
	+ 2\omega_T r^2 d\varphi dt + dz^2.
\end{eqnarray}
	We will use from now on such coordinates $x^{{\mu}}=(t,r,\varphi,z)$. An observer ${\cal O}$ at rest in the rotating system has a four velocity $u^{{\mu}}=(u^{ 0},0,0,0)$, where $u^{ 0}=(1-\omega_T^2 r^2)^{-1/2}$ in order to have $u^{ \mu} u_{ \mu}=-1$. The observed velocity of a particle is given by $v\equiv(v^{ 0},v^r,v^{\varphi},v^z)$, where $v^{ 0}$ can be found by imposing that $v\cdot u = 0 $.

%In the rotating case we may adopt coordinates of the lab,  $u=(u^0,0,0,0)$. 
We will work at linear order in $\omega_T$, {\it i.e.}, considering only Coriolis forces and neglecting centrifugal terms of ${\cal O}(\omega_T^2  R)$, where $R$ is the Earth radius. The latter are indeed smaller, at relativistic velocities, by about $|\omega^2_T R|/|\omega_T\times v|\approx  0.5 \times 10^{-6}$.

By computing the Christoffel symbols from the above metric one finds
\begin{eqnarray}
f_G &\simeq& \omega_T \gamma (0,r v^\varphi , -v^r/r, 0 )= - \gamma (\Omega_T \times_u v ) \,, \label{fGCoriolis}
%f_G\cdot S &=& \Omega \cdot (v \times_u S) \, ,
\end{eqnarray} 
where $\Omega_T \equiv (0,0,0,  \omega_T)$.
The only non-trivial quantity at linear order in $\omega_T$ is a Coriolis-type effect 
\begin{eqnarray}
%f_G \cdot v &=& -r v^r \gamma \gamma_T^2 \omega^2 \\
f_G\cdot S &\simeq& - \gamma \,  \Omega_T \cdot (v \times_u \Sigma) \, .
\end{eqnarray}
%where $\Omega_T \equiv (0,0,0,\omega_T)$.

Using Eq.~(\ref{genericmetric}) and comparing with Eqs.~(\ref{dEdT}),~(\ref{omegaadef}),  we obtain the Coriolis contribution  $\bfomega^{\rm Cor}_a$ (at linear order in $\omega_T$) in analogy  to the anomalous precession vector $\bfomega_a$:
\begin{eqnarray}
 \bfomega^{\rm Cor}_a \equiv - \frac{\bf \Omega_T}{\gamma} \, . \label{Rotation1}
\end{eqnarray}
%where   $\bfomega^{ROT}_a$ is the new precession vector. This amounts to a relative correction of order $ \delta\omega_a/\omega_a\approx \omega_T/(\gamma \omega_a) \approx1.7\cdot 10^{-12}$, at the magic momentum.
(In the Newtonian limit {\it e.g.} see \cite{JacksonKimball:2017elr}.) The same result can be obtained directly using Eq.~(\ref{dE/dT}). Indeed, considering $\Omega_{\rm tot}=\Omega_v - \Omega_\Sigma + \Omega_{\mathcal{E}}$ and making explicit $\Omega_v$, $ \Omega_\Sigma$ and $\Omega_{\mathcal{E}}$, for the case $E=B=0$, at linear order in $\omega_T$ (and thus ignoring terms in $\dot \gamma $, which are quadratic in  $\omega_T$) we recover the above result.

\subsubsection{Equations of motion and numerical solution}

\begin{figure}
\centering
	\includegraphics[height=6.5cm,width=8cm]{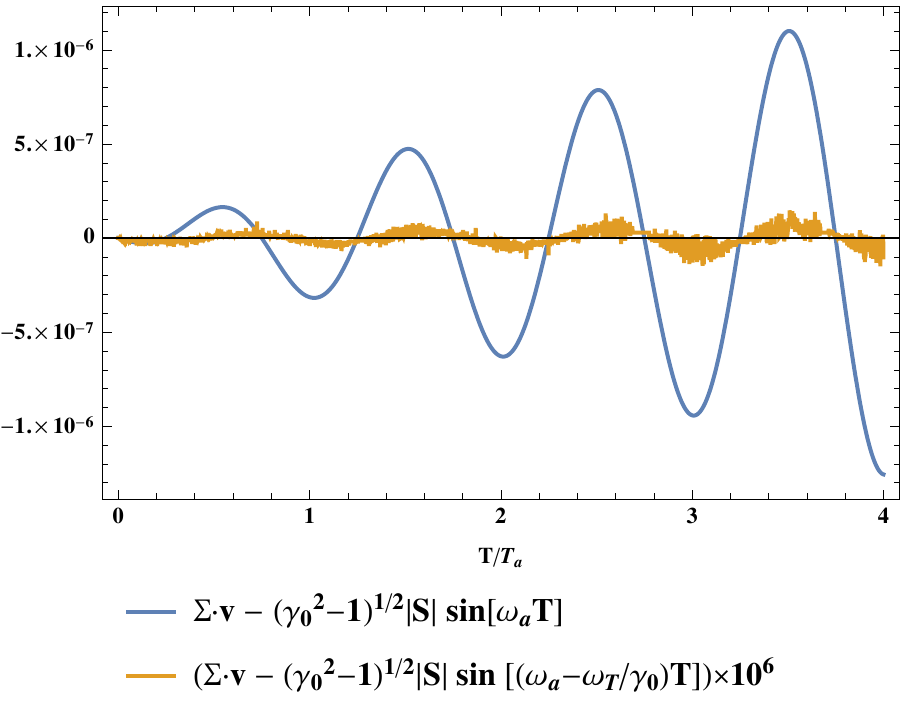} 
%	\,

	\caption{We have  solved numerically the equations of motion Eqs.~(\ref{eqsv})-(\ref{eqsSigma}) and then evaluated $\Sigma\cdot v$, from Eq.~(\ref{Sigmav}), on the solution. We show that the sinusoidal behavior with frequency $\omega_a-\omega_T/\gamma_0$  has the correct behavior, modulo tiny higher order effects. We have set  realistic values of $a$ and $\omega_T$ and an initial condition $\gamma(T=0)\equiv \gamma_0=29.3$. Here $T_a\equiv 2\pi/\omega_a$. Note that also $\gamma$ is constant at linear order in $\omega_T$, and so the only time-dependent part of ${\cal E}$ in Eq.~(\ref{omegaeq}) is due to $\Sigma\cdot v$.}
\label{fig:omegaPvsomega}
%	}
\end{figure}

We  double checked the previous result by solving numerically the exact equations of motion, adding also the presence of a constant magnetic field. The equations of motion become
\begin{eqnarray}
\frac{Dv}{DT} &=&-\frac{v\cdot f_G}{\gamma} (u+v) + \frac{f_G}{\gamma}+ \frac{e}{m \gamma} (v\times_u B) \, , \label{eqsv} 
\end{eqnarray}
and
\begin{eqnarray}
\frac{D\Sigma}{DT} &=&-\frac{\Sigma\cdot f_G}{\gamma} u + (\Sigma\cdot v) \frac{f_G}{\gamma}+\frac{g e}{2 m \gamma} (\Sigma\times_u B) -\frac{e}{m} a  v \gamma \Sigma\cdot (v\times_u B) \, .
\label{eqsSigma}
\end{eqnarray}
We then need to evaluate the quantities $\Sigma\cdot v$ and $\gamma=1/\sqrt{1-v^2}$ inside the electron energy ${\cal E}$, given by Eqs.~(\ref{omegaeq}), (\ref{omegazero}). In the rotating metric we have that:
\begin{eqnarray}
v^2=(v^r)^2+\frac{r^2 (v^{ \varphi})^2}{1-r^2 \omega_T^2}+(v^z)^2 \, ,\qquad \, 
\Sigma\cdot v=v^r \Sigma^r+\frac{r^2 v^{ \varphi} \Sigma^{ \varphi}}{1-r^2 \omega_T^2}+v^z \Sigma^z \, .
\label{Sigmav}
\end{eqnarray}
We only need therefore to solve for the $r,\varphi,z$ components of Eqs.~(\ref{eqsv}) and (\ref{eqsSigma}).
One can work with the  above full equations numerically. We have indeed solved them in the particular case of $B=(0,0,0,|B|)$, where $B$ and the Earth rotation are aligned. The results in Fig.~\ref{fig:omegaPvsomega} show that the leading effect is captured by the vector $ \bfomega^{\rm Cor}_a$ of Eq.~(\ref{Rotation1}). We infer thus that, at linear order in $\omega_T$, the total precession is given by
%Using Eq.~(\ref{genericmetric}) and comparing with Eq.~(\ref{dEdT}) the anomalous precession vector $\omega_a$ gets corrected as follows:
\begin{eqnarray}
 \bfomega_a \rightarrow   \bfomega_a +  \bfomega^{\rm Cor}_a \simeq \frac{e}{m} a {\bf B}-\frac{\bf \Omega_T}{\gamma} \, . \label{Rotation2}
\end{eqnarray}
This amounts to a relative correction of order $ \Delta\omega_a/\omega_a\approx \omega_T/(\gamma \omega_a) \approx1.7\cdot 10^{-12}$, at the magic momentum.
Note that also $\gamma$ can be taken to be constant at linear order in $\omega_T$.

This behavior can be studied also at linear order in $\omega_T$, leading to a more explicit form, which has also been used as a numerical test, giving the same leading order results. The first equation becomes
\begin{eqnarray}
\frac{Dv^i}{DT} &=& \frac{f_G^i}{\gamma}+ \frac{e}{m \gamma} (v\times B)^i \, . \label{eqsvCor} 
\end{eqnarray}
The spin equation becomes instead
\begin{eqnarray}
\frac{D\Sigma^i}{DT} &=& (\Sigma\cdot v) \frac{f_G^{i}}{\gamma}+\frac{g e}{2 m \gamma} (\Sigma\times B)^i -\frac{e}{m} a  v^i \gamma \Sigma\cdot (v\times B) \, .
\label{eqsSigmaCor}
\end{eqnarray}
In both equations all the scalar and vector products do {\it not} contain $\omega_T$ and the linearized $f_G$ is the only term proportional to $\omega_T$,  given by Eq.~(\ref{fGCoriolis}).
The covariant derivatives also contains terms proportional to $\omega_T$, since:
\begin{eqnarray}
\frac{Dv^r}{DT} &=& \frac{d v^r}{dT}- r (v^{ \varphi})^2 - r v^{ \varphi} \omega_T \, , \nonumber \\ 
\frac{Dv^{ \varphi}}{DT} &=& \frac{d v^{ \varphi}}{dT}+ \frac{2 v^r v^{ \varphi} }{r} + \omega_T \frac{v^r}{r} \, , \ \nonumber \\ 
\frac{Dv^z}{DT} &=&  \frac{d v^z}{dT} \, , \nonumber \\ 
\frac{D\Sigma^r}{DT} &=& \frac{d \Sigma^r}{dT}- r v^{ \varphi} \Sigma^{ \varphi} - r \Sigma^{ \varphi} \omega_T \, , \ \nonumber \\ 
\frac{D\Sigma^{ \varphi} }{DT} &=& \frac{d \Sigma^{ \varphi} }{dT}+ \frac{\Sigma^{ \varphi}  v^r + \Sigma^r v^{ \varphi} }{r} + \omega_T \frac{\Sigma^r}{r} \, , \nonumber \\ 
\frac{D\Sigma^z}{DT} &=&  \frac{d \Sigma^z}{dT} \, .
\end{eqnarray}

As a final comment note that, while the Coriolis effect is very small compared to the experimental sensitivity for muon $g-2$ experiments, it could be slightly more important for {\it electron} $g-2$ experiments, which are much more precise. In the electron experiments velocities are non-relativistic and so the Coriolis frequency vector is just ${\bf \Omega}_T$. This should be compared with the Larmor frequency of about $\omega_L=100 \, {\rm GHz}$, for an electron in a Penning trap experiment~\cite{Hanneke:2008tm,Sturm:2013upa}, which gives a relative shift $\Delta\omega/\omega_L\approx 7\times 10^{-16}$, to be compared with the experimental relative error on $g$, at present of order ${\cal O}(10^{-13})$, and forecasted to reach ${\cal O}(10^{-14})$~\cite{Gabrielse:2019cgf} within one or two years from now. In terms of $a_{\rm el}\equiv g/2-1$ the effect amounts to a relative shift $\Delta  a_{{\rm el}}/a_{{\rm el}}  \approx 7\times 10^{-13}$, compared to the  experimental relative error  $\Delta  a_{{\rm el}}/a_{{\rm el}}  \approx {\cal O}(10^{-10})$.

\subsubsection{Schwarzschild metric}

In this subsection we consider the even more negligible effect of Earth's gravity.
As argued above  by a rough power counting we expect indeed a suppression of about three orders of magnitude compared to the Coriolis effect.
	
	We consider here a standard Schwarzschild metric, given by
	\begin{eqnarray}
	ds^2=-dt^2 \left(1-2 \frac{GM}{r} \right)+\frac{dr^2}{1-2\frac{GM}{r}}+r^2d \theta^2 + r^2 \sin(\theta)d\varphi^2 \, .
\end{eqnarray}
In this case for an observer at fixed coordinates, with $u=(u^0,0,0,0)$, we get:
\begin{eqnarray}
f_G=\left(0,-\frac{\gamma  GM}{r^2},0,0\right) \, ,	
\end{eqnarray}
leading to
\begin{eqnarray}
f_G\cdot v= \frac{\gamma  GM v^r}{2 GM r-r^2}  \,  , \qquad \, 	f_G\cdot S &=& \frac{ \gamma G M  S^r}{2 G M r-r^2} \, , \label{fGvandfGS}
\end{eqnarray}
where $S=(S^0,S^r,S^\theta,S^\varphi)$.
At leading order in $GM/r$ we get
\begin{eqnarray}
%f_G \cdot v &=& -r v^r \gamma \gamma_T^2 \omega^2 \\
f_G\cdot v= - g_T \gamma v^r  \, , \qquad
f_G\cdot S &=& - g_T \gamma S^r \, ,
\end{eqnarray}
where at the Earth's surface $g_T\equiv G M / r^2$.
The first effect is simply due to the radial acceleration of a free falling muon and it is thus suppressed by the smallness of $g_T$ and furthermore by the smallness of the radial velocities, $v_r\ll 1$, in a realistic experiment and so, comparing Eq.~(\ref{genericmetric})  with Eq.~(\ref{dEdT}), it  leads to a relative error $\Delta\omega_a/\omega_a\approx g_T v^r/(\omega_a) \approx 10^{-15} \, v^r$ and thus totally negligible. The second effect is less suppressed and comparing again with Eq.~(\ref{dEdT}) one gets $\Delta\omega_a/\omega_a\approx g_T/(\gamma \omega_a) \approx 8\times 10^{-16}$, taking into account that, using Eq.~(\ref{Slab}), one has $S^r\approx s^r$ in a realistic experiment,  since $v^r\ll1$.

One can further check this more explicitly by writing a system of equations in a simple setup, where the magnetic field is along $r$, $B=(0,B^r,0,0)$. In this case the observed magnitude of the magnetic field is given by
\begin{eqnarray}
|B|=\sqrt{B\cdot B}=B^r/\sqrt{1-2\Phi} \, , \qquad {\rm where} \qquad    \Phi\equiv g_T r  \, .\label{modB}
\end{eqnarray}
Here we see that if we use the system of coordinates of the observer $u$, then $B^r/\sqrt{1-2\Phi} = B^{\hat r}$.
Moreover in a realistic laboratory setup  we can also neglect terms in $v^r$.
The explicit equations of motion Eqs.~(\ref{eqsv}) and (\ref{eqsSigma}) get  additional terms only in the two following equations 
\begin{eqnarray}
\frac{d v^r}{dT}&=&\frac{d v^r}{dT}\Big|_{\rm flat space} -g_T \,  ,\\
\frac{d \Sigma^r}{dT}&=&\frac{d \Sigma^r}{dT}\Big|_{\rm flat space} -3 g_T  r^2 (s^\theta v^\theta + s^\varphi v^\varphi \sin^2\theta ) \, ,
\end{eqnarray}
where $d/dT|_{\rm flat space} $ contain the usual flat space terms, with $|B|$ defined in the above Eq.~(\ref{modB}).
The solution may be plugged then into
\begin{eqnarray}
\Sigma\cdot v = r^2 (s^\theta v^\theta + s^\varphi v^\varphi \sin^2\theta ) \, ,
\end{eqnarray}
while terms in $\dot{\gamma}$ are neglected.
By comparing with Eqs.~(\ref{eqsvCor}) and (\ref{eqsSigmaCor})  an order of magnitude estimate of the effects is obtained, for relativistic velocities, simply by replacing $\omega_T$ with $g_T$, in agreement with the above $\Delta\omega_a/\omega_a\approx g_T/(\gamma \omega_a) \approx 8\times 10^{-16}$.

As a side comment, we note that our formalism might be useful to study the case of a spinning body close to a system with strong gravity, such as a black hole. From eq.~(\ref{fGvandfGS}) one can see that the effects become very large close to the horizon, when $r$ approaches the Schwarzschild radius, $r_s\equiv 2 G M$.

\subsubsection{Gravitational wave}

We also discuss here, as a very simple application of our formalism, the case of a gravitational wave (GW) metric. This is negligible for a realistic $g-2$ experiment, due to the smallness of the GW's amplitude on Earth, but it might have applications in other setups, such as the case of a spinning body close to a strong source of GW's, {\it e.g.} a black hole or neutron star binary system.
We may consider thus a GW metric, as a perturbation of flat spacetime in Cartesian coordinates, using Lorenz and Transverse Traceless gauge conditions, which completely fix the coordinates. For instance the metric for a GW propagating in the $\hat{z}$ direction with $+$ and $\times$ polarizations read
\begin{eqnarray}
	ds^2=-dt^2 +dx^2(1+h_+(t,z)) +dy^2(1+h_+(t,z)) + 2 h_\times dx dy +dz^2  \, .
	\end{eqnarray}
Assuming an observer at rest $u=(1,0,0,0)$, one finds, at linear order in $h_+$ and $h_\times$,
\begin{eqnarray}
f_G= -\frac{\gamma}{2} \left(0,v^x \frac{dh_+}{dt}  + v^y  \frac{dh_\times}{dt}  ,   v^x  \frac{dh_\times}{dt} - v^y \frac{dh_+}{dt}  ,0\right) \, ,	
\end{eqnarray}
leading to
\begin{eqnarray}
f_G\cdot v &=& \gamma \left( \frac{dh_+}{dt}  \frac{(v^y)^2-(v^x)^2}{2} - \frac{dh_\times}{dt} v^x v^y    \right) \,  , \nonumber \\	f_G\cdot S &=& \frac{\gamma}{2} \left( \frac{dh_+}{dt} (S^y v^y-S^x v^x) -  \frac{dh_\times}{dt}  (S^y v^x +S^x v^y)  \right)  \, , 
\end{eqnarray}
where $S=(S^0,S^x,S^y,S^z)$ and  $v=(v^0,v^x,v^y,v^z)$.
For a gravitational wave of amplitude $h$ and angular frequency $\omega_{\rm GW}$, the effects on a relativistic particle are of order $f_G\cdot v \approx f_G\cdot S/\gamma \approx \gamma h \omega_{\rm GW}$, using also Eq.~(\ref{Slab}). Comparing again Eq.~(\ref{genericmetric})  with Eq.~(\ref{dEdT}), these represent corrections to the precession frequency of order $\Delta\omega_a/\omega_a\approx h \omega_{GW}/\omega_a$. For GW's detected on Earth, such as the ones coming from coalescing binaries, with $h\sim 10^{-21}$ and $\omega_{GW}\sim 10^2 {\rm Hz}$, this is about $10^{-25}$.

%\frac{e}{m} E
\section{Impact on Experiments}\label{sec:Experiments}

\subsection{Possible systematic effects due to $E$}

From the previous sections we have found that the leading effect of a non-trivial metric, the Coriolis force, leads to a shift  $\Delta\omega/\omega_a \approx 10^{-12}$, much smaller than the present sensitivity. However we show here that effects purely due to electromagnetism have to be taken into account: 
\begin{itemize}
\item[(1)] a proper treatment of the ``pitch'' correction; 
\item[(2)] a proper treatment of the presence of electric fields; 
\item[(2)] the fact that the muon $\gamma$ factor is not constant, also because of electric fields, which has given rise to the first two terms in Eq.~(\ref{eqE}) that do not vanish at the magic momentum and might not be small in present experiments.
\end{itemize}
We  estimate carefully such effects here, using a Minkowski metric, and considering a realistic experimental setup. 

The experimentally observable quantity in Eqs.~(\ref{energyel}) and (\ref{omegaeq}) is 
\begin{eqnarray}
\mathcal{E}=\mathcal{E}_0 + \mathcal{E}_R=m_e \gamma_e^{\rm (M)} \gamma-\mathcal{E}_1 \gamma |v| |S| (\hat{s}\cdot \hat{v}) =m_e \gamma_e^{\rm (M)}\gamma-\mathcal{E}_1 |S| \sqrt{\gamma^2-1}  (\hat{s}\cdot \hat{v})  \, . \label{Eelectron}
\end{eqnarray}

%We study now the impact of such oscillations on Eq.~(\ref{Eelectron}), namely on the time dependence of the precession, described by $\hat{s}\cdot \hat{v}$, and of the energy $\gamma$.
As we will see, in a realistic experiment there are non-trivial effects both in $\hat{s}\cdot\hat{v}$, and in $\gamma$, leading to deviations from the simple sinusoidal time dependence with frequency $\omega_a$.

We follow here Ref.~\cite{Miller:2018jum}.
At zero order we consider a perfectly circular motion, with constant gamma factor, chosen to be equal to the ``magic" gamma factor, $\gamma_0=1/\sqrt{1-v_0^2}=\gamma_M$,  in a constant magnetic field of magnitude $|B|$, at the cyclotron frequency
\begin{eqnarray}
\omega_c \equiv \frac{e}{m}  \frac{|B|}{\gamma_M} \label{eqleading} \, ,
\end{eqnarray}
and at the equilibrium radius
\begin{eqnarray}
R_0= \frac{m}{e} \frac{\gamma_M v_0}{|B|} \, .
\end{eqnarray}
Eq.~(\ref{Eelectron}) becomes
\begin{eqnarray}
 \mathcal{E} &=&  \mathcal{E}_0 -  \mathcal{E}_1 \gamma_0 |v_0| |S|   \sin(\omega_a T) \, \label{eqleading} \, ,
\end{eqnarray}
with constant $ \mathcal{E}_0 $ and where we have used as an initial condition a spin orthogonal to both $B$ and the velocity.

We evaluate then the size of the corrections relative to the above leading behavior.
We model the electric field from a quadrupole potential in cylindrical coordinates ($r$, $\varphi$ and $z$ are respectively the radial, azimutal and vertical coordinate):
\begin{eqnarray}
V(r,z) = -\frac{1}{2} \kappa \left( (r-R_0)^2-z^2 \right) \, .
\end{eqnarray}
This leads to the following electric field:
\begin{eqnarray}
{\bf E}=- \kappa z {\bf \hat{z}} + \kappa (r-R_0) {\bf \hat{r}} \, .
\end{eqnarray}

A muon can be displaced from the horizontal plane or it can be displaced radially from the equilibrium orbit, leading respectively to vertical or radial oscillations. We analyze them separately.

\subsubsection{Vertical oscillations}
A muon displaced from the horizontal plane will perform small oscillations due to the electric field in the $\hat{z}$ direction (see Appendix III in Ref. \cite{Miller:2018jum}),
%\begin{eqnarray}
%{\bf v}=|v| (\cos\psi \,  \varphi/R_0 + \sin\psi \, \hat{z}) , \, \qquad \psi=\psi_0 \cos(\omega_z t) \, ,
%\end{eqnarray}
%where $\psi_0\ll 1$, so that approximately $v_z \approx |v| \psi_0 \cos(\omega_z t) $. The position of the muon along the $\hat{z}$ direction is thus given by:
%\begin{eqnarray}
%z(t) \approx \psi_0/\omega_z \sin(\omega_z t) \, .
%\end{eqnarray}
\begin{eqnarray}
z(T) &=& A_z \cos(\omega_z T +\varphi_{0,z}) \, ,  \nonumber \\
v_z(T) &=& -A_z \omega_z \sin(\omega_z T +\varphi_{0,z}) , \label{zetat}
\end{eqnarray}
where 
\begin{eqnarray}
\omega_z=\omega_c\sqrt{n} \, , \qquad {\rm with} \,  \qquad n\equiv \frac{\kappa R_0}{|v| |B|} \, .
\end{eqnarray}
We have first checked such behavior by numerically solving the differential equations for velocity and spin using the following values,
 appropriate for the experiment~\cite{Bennett:2006fi}:
\begin{eqnarray}
%\nu_x &\approx&0.93 \, ,  \\
A_z &=&  22.5 \, {\rm mm} \, ,\\
%R_0 &\approx& 7.112 m \, ,\\
\kappa &=&2\cdot 10^7 \frac{{\rm V}}{{\rm m}^2} \, ,\\
B &=& 1.5 {\rm T} \, , \\
a &=& 0.00116585 \, , \label{numvalues}
\end{eqnarray}
which implies $R_0 \approx 7.112$ m and $\omega_a = 2\pi\times 0.23 \, {\rm MHz}$. We have set also, as an initial condition, the magic momentum $\gamma_0 = \gamma_M\equiv \sqrt{\frac{1+a}{a}} \approx 29.3$  and we have considered an initial displacement along the $\hat{z}$ axis equal to $A_z$. The value for $A_z$ has been chosen using half of the radius of the storage volume, $r_{\rm max}=45 {\rm mm}$.

Our numerical results reproduce the above vertical oscillations, except for a small frequency shift $\omega_z\rightarrow \omega_z+\delta\omega_z$, where $\delta\omega_z\approx 2.9072 \times 10^{-4} \omega_z$. 

We study now the impact of such oscillations on Eq.~(\ref{Eelectron}), namely on the time dependence of the precession, described by $\hat{s}\cdot \hat{v}$, and of the energy $\gamma$.

The vertical oscillations are known to modify the precession frequency by the so-called ``pitch correction''. This is usually evaluated by expanding at lowest order Eq.~(\ref{omegaPsquared}) and then taking a time average $\langle ... \rangle$ of its modulus over the fast vertical oscillations. This would lead (modulo a phase) to
\begin{eqnarray}
\hat{s}\cdot \hat{v} = \sin(\tilde{\omega}_a T) \, , \, \, \, \, \,  {\rm with} \, \, \, \, \tilde{\omega}_a\equiv \langle |\tilde{\bfomega}_a| \rangle \approx \omega_a - \left\langle \frac{({\bf v}\cdot {\bf \hat B})^2}{2} \right\rangle  = \omega_a \left(1 - \frac{(A_z \omega_z)^2}{4} \right) \equiv \omega_a+\delta\omega_P\, . \nonumber \\ \label{omegaP}
\end{eqnarray}
With the above values $\delta\omega_P\approx -8\cdot 10^{-7} \omega_a$, which is taken into account by the experimental collaborations~\cite{Bennett:2006fi,Grange:2015fou}. However we point out that the true frequency should be given instead by Eqs.~(\ref{secondderivative}) and (\ref{finalomega2}), which depends also on the time evolution of the spin $s^z(T)$.

We have first checked the true value of the ``pitch correction'' numerically. We find that $\delta\omega_P$ gives a reasonably good  approximation, but we also find a subleading correction $\delta\omega_{SL}$, so that the true frequency is
\begin{eqnarray}
\omega_T=\tilde{\omega}_a+\delta\omega_{SL} \, ,\qquad \delta\omega_{SL} \approx 1.75 \cdot 10^{-3} \delta\omega_P \, . \label{SL}
\end{eqnarray}
This is shown in Fig.~\ref{pitch}.

\begin{figure}
  \begin{tabular}{p{0.48\textwidth} p{0.48\textwidth}}
  % p{0.3\textwidth}}
 \vspace{0pt}  \includegraphics[width=0.48 \textwidth]{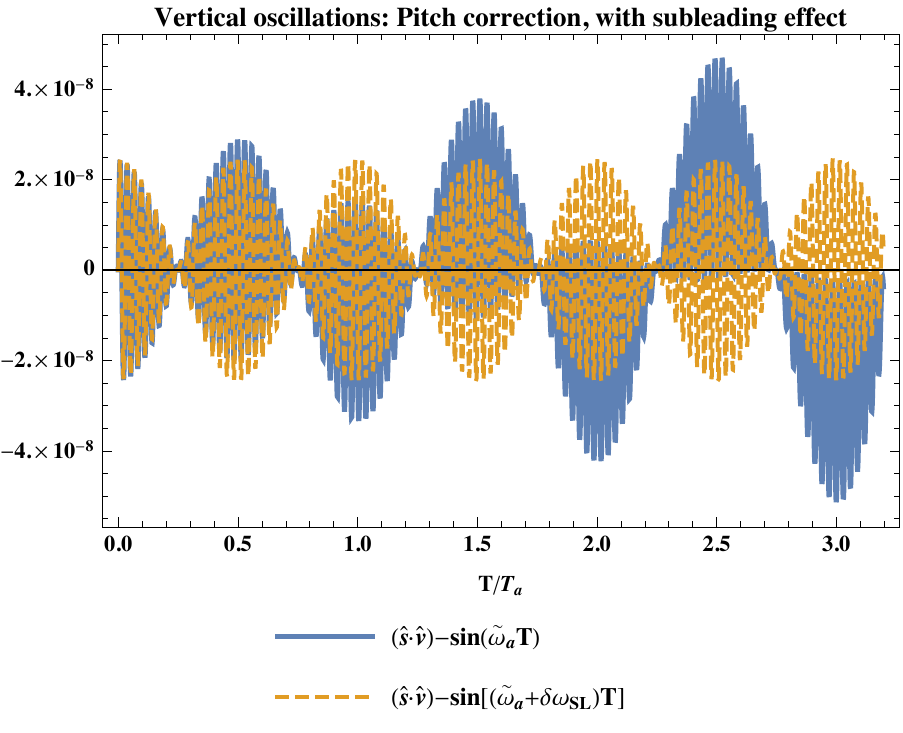} &
 \vspace{-3pt}         \includegraphics[width=0.48 \textwidth]{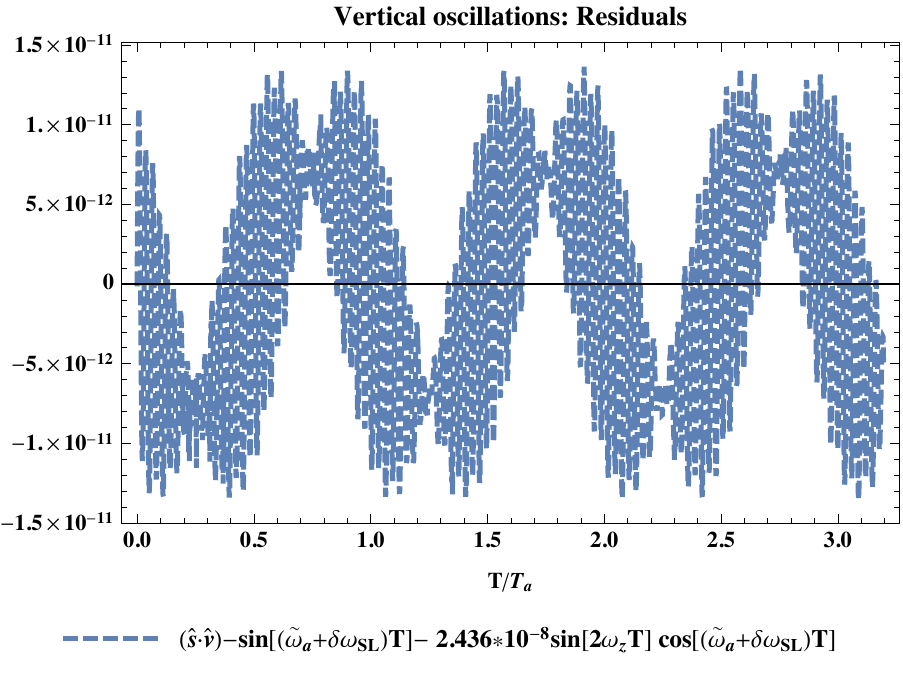} 
%%    \vspace{-1pt}        \includegraphics[width=0.3 \textwidth]{Vdependence.pdf} 
    \end{tabular}  %  
%	\centering
%	\includegraphics[height=5.8cm]{pitch} 
%	%\vspace{-1cm}
%	\includegraphics[height=5.3cm]{pitchresidual} 
%%	\,
	\caption{We have  solved numerically the equations of motion and then evaluated $\hat{s}\cdot \hat{v}=(\Sigma \cdot v)/(\gamma |v| |S|)$ on the solution. The left plot shows that a better fit is achieved by adding the subleading correction $\delta\omega_{SL}$ of Eq.~(\ref{SL}) to the traditional ``pitch correction'', contained in $\tilde{\omega}_a$. Here $T_a\equiv 2\pi/\omega_a$. The residual is well fit (right plot) by the functional form $\epsilon \sin(2\omega_z T) \cos((\tilde{\omega}_a+\delta\omega_{SL}))$, with $\epsilon=2.436\times10^{-8}$.}
\label{pitch}
%	}
\end{figure}

\begin{figure}
	\includegraphics[height=7cm,width=15cm]{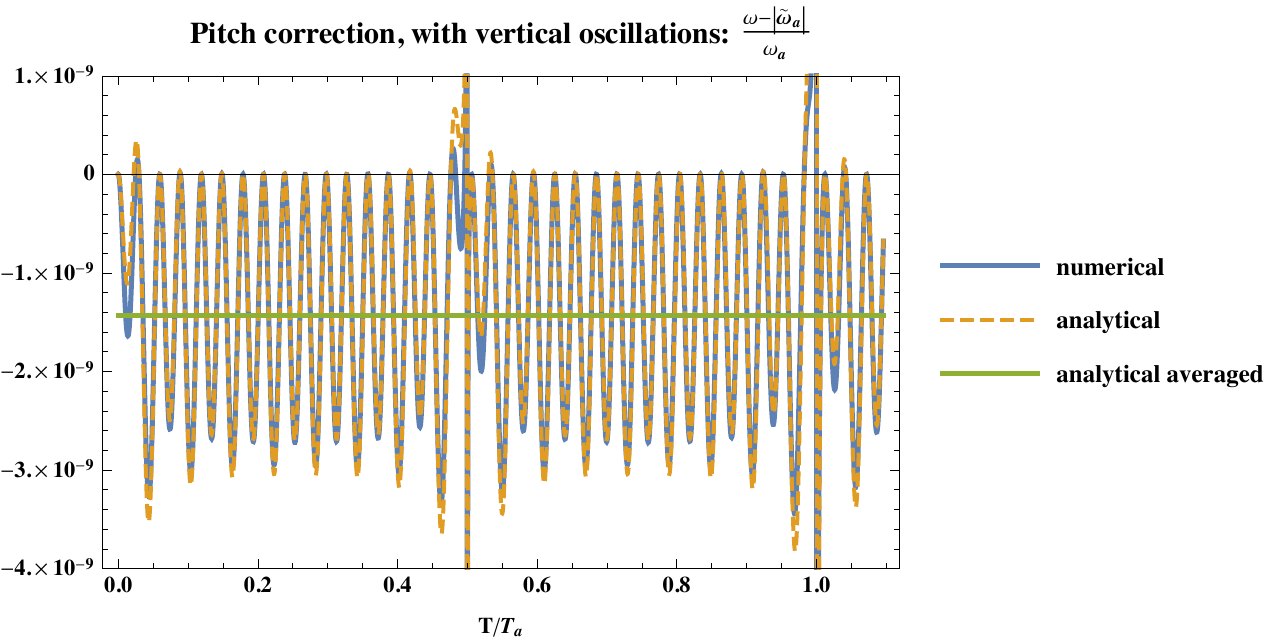} 
%	\,
   \captionsetup{singlelinecheck=off}
\caption[.]{We have  solved numerically the equations of motion and then evaluated \[
      \omega\equiv \omega_a \left[ 1- \frac{(\hat B\cdot \Sigma)(\hat \Sigma \cdot \hat B))}{(\hat \Sigma\cdot v)}\right]^{1/2}\;,
    \]
%	$ \omega\equiv \omega_a \left( 1- \frac{(\hat B\cdot \Sigma)(\hat \Sigma \cdot \hat B))}{(\hat \Sigma\cdot v)}\right)^{1/2}$
from Eq.~(\ref{secondderivative}) and $|\tilde{\bfomega}_a|$ (which contains the traditional ``pitch'' correction) from Eq.~(\ref{omegaPsquared}), on the solution. We also show the very good agreeement with our analytical result of Eq.~(\ref{analyticalpitch}). Taking a time average over the fast oscillations we get precisely the subleading correction $\delta\omega_{SL}$ of Eq.~(\ref{SL}), which also coincides with Eq.~(\ref{analyticalmean}) (green line). Here $T_a\equiv 2\pi/\omega_a$ and we used $\Sigma^z_0=0$.}
\label{omegaPvsomega} 
%	}
\end{figure}

It turns out that the quantity $\omega$ defined in our Eqs.~(\ref{secondderivative}) and (\ref{finalomega2}) capture such additional terms since we find that $\langle \omega \rangle=\tilde{\omega}_a+ \delta\omega_{SL}$, with very good precision. This can be checked numerically, evaluating Eqs.~(\ref{secondderivative}) and (\ref{finalomega2}) on the full solutions, as can be seen in fig.~\ref{omegaPvsomega}. We can also evaluate it analytically, solving the differential equations for $\Sigma^z$. The full equation is
\begin{eqnarray}
\frac{d \Sigma^z}{dT}= \frac{e}{m}\left[ \frac{1+a}{\gamma}E^z(\Sigma\cdot v) - a \gamma v^z  \left( B\cdot(\Sigma\times v) + (\Sigma\cdot E) - (v\cdot \Sigma) (E\cdot v) \right) \right] \, ,
\end{eqnarray}
but we have checked numerically that at leading order we may: {\it i}) neglect the last two terms,  {\it ii}) consider $\gamma=\gamma_0$, {\it iii}) use the zeroth order solution for $\Sigma\cdot v=\gamma_0 v_0 |S| \sin(\omega_a T)$ in the r.h.s,  {\it iv}) use Eqs.~(\ref{Sigmas2}) and (\ref{omegavec2}) for $B\cdot(\Sigma\times v)$. 
This leads to a simpler equation 
\begin{eqnarray}
\frac{d \Sigma^z}{dT}=v_0 |S| \left[\omega_a \gamma_0   v^z \cos (\omega_a t)+(1+a) \frac{e}{m}\kappa z(t) \sin (\omega_a t) \right]\, .
\end{eqnarray}
One can now use now Eqs.~(\ref{zetat})  and integrate in time, finding
\begin{eqnarray}
\Sigma^z(T) &=& -A_z\frac{ \omega_z \sin (T \omega_a) \sin (T \omega_z) \left[(a+1) e \kappa +\gamma_0 m   \omega_a^2\right]+\omega_a \cos (T \omega_a) \cos (T \omega_z) \left[(a+1) e \kappa +\gamma_0 m   \omega_z^2\right]}{m \left(\omega_a^2-\omega_z^2\right)}   \nonumber \\
& & + A_z \frac{\omega_a \left[(a+1) e \kappa +\gamma_0 m \omega_z^2\right]}{m \left(\omega_a^2-\omega_z^2\right) } +\Sigma_0^z \, ,
\end{eqnarray}
where we used the initial condition $\Sigma^z(t=0)=\Sigma_0^z$. 
As a final step we plug this into Eq.~(\ref{secondderivative}) and compare with Eq.~(\ref{omegaPsquared}), leading, at linear order, to
\begin{eqnarray}
\frac{\omega-\tilde{\omega}_a}{\omega_a}&=&\frac{A_z^2}{\gamma_0 {\omega}_z}  \sin (\omega_z  T) \bigg[\frac{e}{m}\kappa(1+a)+\gamma_0 {\omega}_z^2\bigg]  \nonumber \\
&&\times \left[\omega_a \cos ({\omega}_z T) \cot (\omega_a T )+\tilde{\omega}_z \sin ({\omega}_z T)-\omega_a \csc (\omega_a T)\right] - \Sigma^z_0 \frac{A_z \omega_z \csc (\omega_a t ) \sin (\omega_z t)}{\gamma_0}\;,\nonumber\\ \, \label{analyticalpitch}
\end{eqnarray}
we have set for simplicity $\varphi_{0, z}=0$ and we have also used the approximations $\omega_a \ll {\omega}_z$ and $v_0\approx 1$. 
As can be seen from figure~\ref{omegaPvsomega} this formula reproduces in a very accurate way the numerical behavior of $\omega(T)$, obtained evaluating Eq.~(\ref{secondderivative}) on the full numerical solution.
Performing now an average over the fast vertical oscillations only the term quadratic in $\sin({\omega}_z T)$ survives, leading to:
\begin{eqnarray}
\frac{\langle\omega\rangle-\tilde{\omega}_a}{\omega_a}=\frac{A_z^2 \left[\frac{e}{m}\kappa(1+a)+\gamma_0 {\omega}_z^2\right]}{4\gamma_0 } \, . \label{analyticalmean}
\end{eqnarray}
This formula reproduces with excellent approximation the subleading correction $\delta\omega_{SL}$.

We may then look at the behavior of the $\gamma$ factor. Using Eq.~(\ref{gammadot}) and, for simplicity $\varphi_{0, z}=0$ ,we get
\begin{eqnarray}
\dot{\gamma}= \frac{e}{m} E_z v_z =   \frac{e}{2m}  \kappa A_z^2  \omega_z \sin(2 \omega_z T ) \, \implies \gamma = \gamma_0 +  \frac{e}{2m}  \kappa A_z^2  \sin^2(\omega_z T)  \, , \label{eqgammaz}
\end{eqnarray}
where we have used the initial condition $\gamma(T=0)=\gamma_0$.

\begin{figure}
  \begin{tabular}{p{0.45\textwidth} p{0.45\textwidth} p{0.45\textwidth}}
  % p{0.3\textwidth}}
 \vspace{0pt}  \includegraphics[width=0.44 \textwidth]{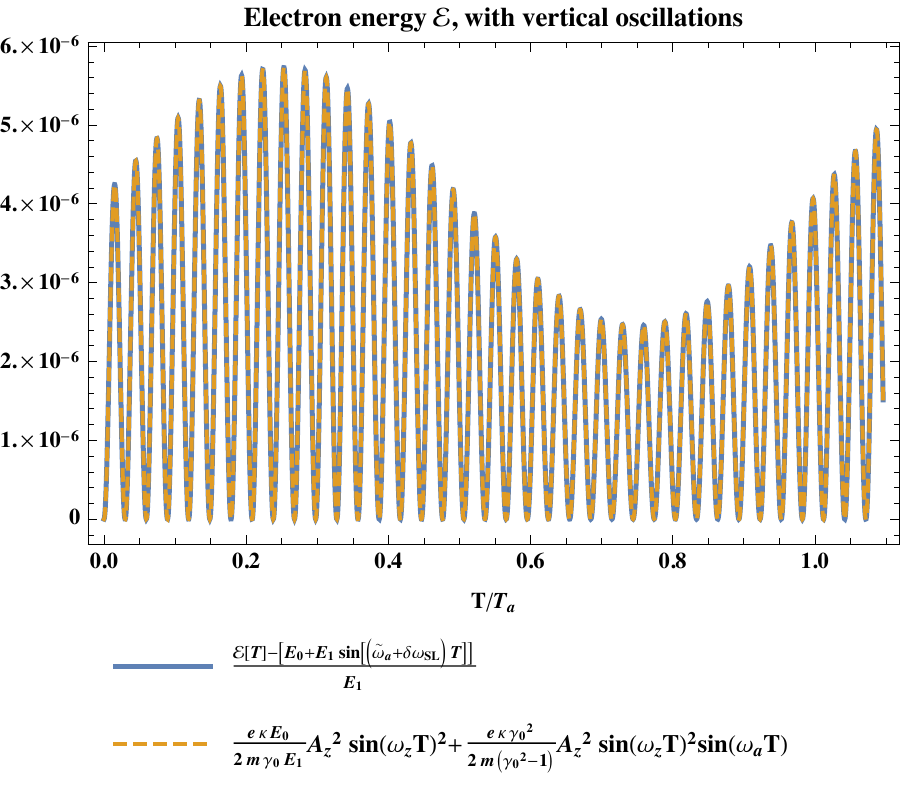} & 
  \vspace{2pt}         \includegraphics[width=0.45 \textwidth]{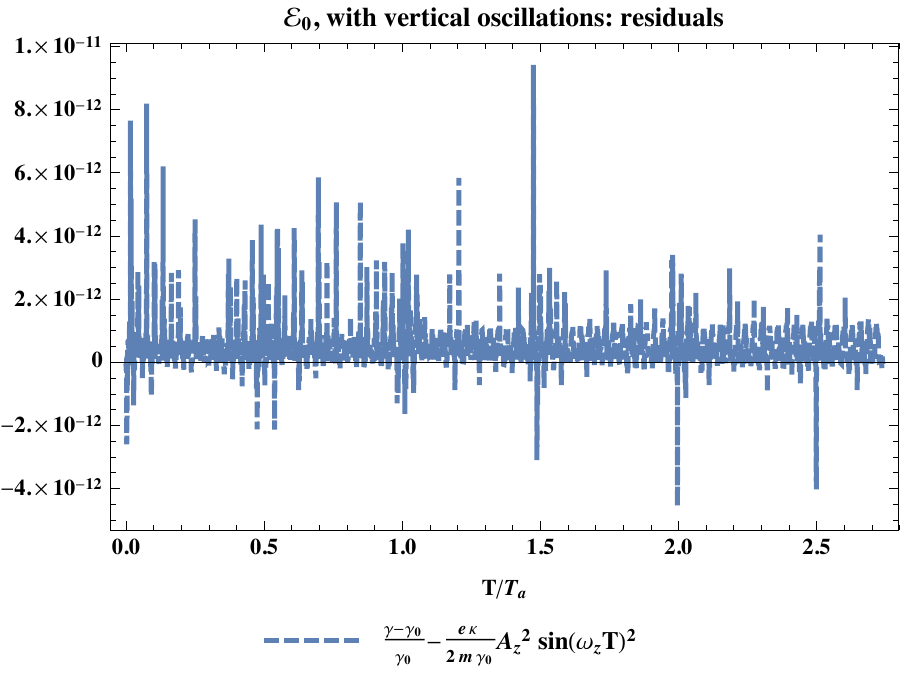} & \\
    \end{tabular} 
        %\vspace{2pt}       
  \begin{center} 
         \includegraphics[width=0.53 \textwidth]{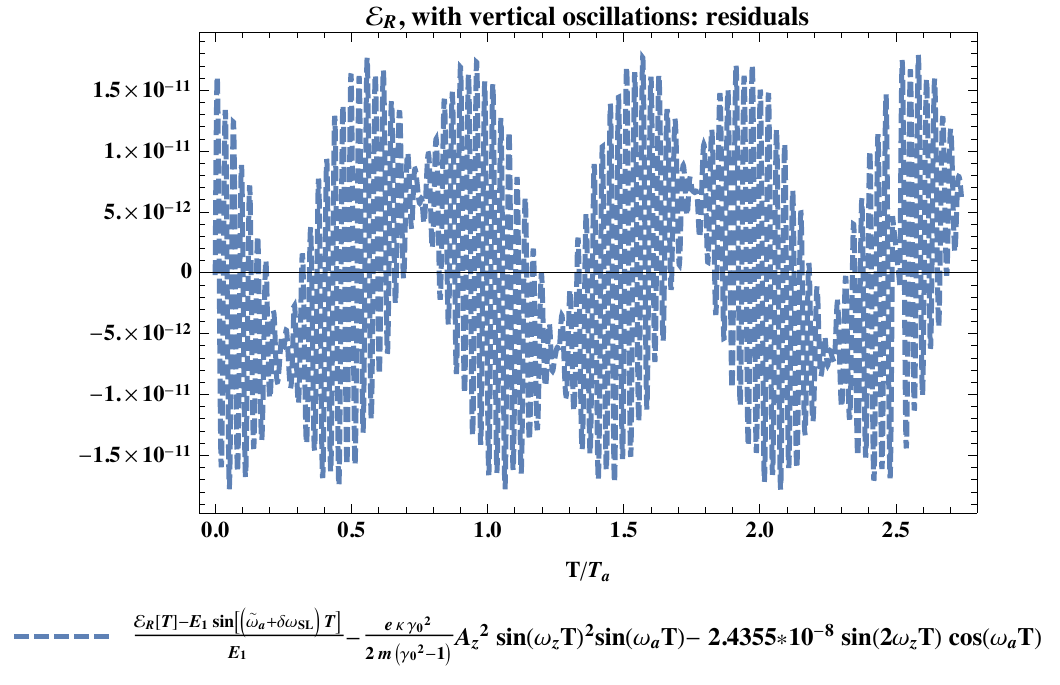} 
 \end{center}
        %&
%    \vspace{-1pt}        \includegraphics[width=0.3 \textwidth]{Vdependence.pdf} 
   % \end{tabular}  %  
%	\includegraphics[height=6.3cm]{energyvert} 
%	\includegraphics[height=5.3cm]{energyvertresidual} 
%	\,

	\caption{We have  solved numerically the equations of motion, with a vertical displacement, and then evaluated the energy ${\cal E}$ on the solution and compared to a simple sinusoidal behavior $E_0+E_1 \sin((\tilde{\omega}_a+\delta\omega_{SL}) T)$, where $E_0$ and $E_1$ are constants. The upper left plot shows the deviation form such a simple behavior and the analytical form of the residual of Eqs.~(\ref{vertresidual0},\ref{vertresidualR}). The upper right plot shows the residual for ${\cal E}_0$, Eq.~(\ref{radialresidual0}). The bottom plot shows an analytical fit of the residual, for ${\cal E}_R$, given by Eq.~(\ref{radialresidualR}),  plus higher order corrections that we have found numerically. Here $T_a\equiv 2\pi/\omega_a$.}		  	\label{fig:energyvert}
%	}
\end{figure}

This leads to the following correction for the ${\cal E}_0$ term of the electron energy in Eq.~(\ref{Eelectron}),
\begin{eqnarray}
\frac{\Delta{\cal E}_0}{{\cal E}_0} \equiv \frac{ \gamma - \gamma_0}{ \gamma_0} =   \frac{e}{2 m \gamma_0}  \kappa A_z^2  \sin^2(\omega_z T) \approx 1.6 \cdot 10^{-6} \ \sin^2(\omega_z T) \, . \label{vertresidual0}
\end{eqnarray}
The ${\cal E}_R$ term is corrected in the following way
\begin{eqnarray}
\frac{ {\Delta \cal { E}}_R}{ {\cal E}_1 |S| \sqrt{\gamma_0^2-1} } &=&\frac{ \sqrt{\gamma^2-1}  (\hat{s}\cdot \hat{v}) - \sqrt{\gamma_0^2-1} (\hat{s}\cdot \hat{v}) }{ \sqrt{\gamma_0^2-1}}  \approx  \frac{e \gamma_0}{2 m (\gamma^2_0-1)}  \kappa A_z^2  \sin^2(\omega_z T)  \sin(\omega_T T) \nonumber \\ &\approx & 1.6 \cdot 10^{-6} \ \sin^2(\omega_z T)  \sin(\omega_T T) \, .
\label{vertresidualR}
\end{eqnarray}
The comparison with numerical results is shown in fig.~\ref{fig:energyvert}.

\subsubsection{Radial oscillations}

We  consider here the radial motion assuming that each muon has an equilibrium radius $R_0+x_e$, where $x_e$ depends on the muon average gamma factor, $\gamma_0$; note that here in general $\gamma_0$ is not equal to $\gamma_M$. If a muon is displaced from this orbit it will perform small radial oscillations, due to the combined effect of the centrifugal term and the radial electric field~\footnote{See Eq.~(15) and (16) in Ref.~\cite{Bennett:2006fi}.}:
\begin{eqnarray}
r(T)=R_0+x_e+A_r \cos(\omega_r T + \varphi_{0,r}) ,\, \qquad  {\rm where} \, \qquad \omega_r \equiv \omega_c \sqrt{1-n} \, ,
\end{eqnarray}
so that:
\begin{eqnarray}
v_r &=& - \omega_r   A_r \sin(\omega_r T + \varphi_{0,r})\, , \\
%\end{eqnarray}
%The radial Electric field is
%\begin{eqnarray}
E_r &=& \kappa [A_r \cos(\omega_r T + \varphi_{0,r})+x_e] \, .
\end{eqnarray}
We will use the same numerical values given above in Eq.~(\ref{numvalues}), with $A_r$ of the same size of $A_z$.

The gamma factor is  determined by
\begin{eqnarray}
\dot{\gamma}= \frac{e}{m} E_r v_r =   -\frac{e}{2m}  \kappa A_r \omega_r [ A_r  \sin(2 \omega_r T ) + 2 x_e \sin(\omega_r T)  ] \, ,
\end{eqnarray}
where $ \varphi_{0,r}$ has been taken equal to zero for simplicity. This implies
\begin{eqnarray}
 \gamma = \gamma_0 -  \frac{e}{2m} \kappa A_r  \left[  A_r  \sin^2 \left(\omega_r T \right) +  4 x_e \sin^2\left(\frac{\omega_r T}{2} \right) \right] \, . \label{eqgammaz}
\end{eqnarray}
This leads to the following corrections:
\begin{eqnarray}
\frac{\Delta{\cal E}_0}{{\cal E}_0}  &=&  - \frac{e}{2 m \gamma_0} \kappa  A_r\left[  A_r  \sin^2(\omega_r T) +  4 x_e \sin^2 \left(\frac{\omega_r T}{2} \right) \right] \nonumber \\ &\approx &- 1.6 \cdot 10^{-6}  \left[\sin^2 \left(\omega_r T\right) + \left( \frac{x_e}{5.625 {\rm mm}} \right)  \sin^2\left(\frac{\omega_r T}{2}\right) \right] \, , 
\label{radialresidual0}
\end{eqnarray}
and
\begin{eqnarray}
\frac{\Delta  {\cal  E}_R}{ {\cal E}_1 |S| \sqrt{\gamma_0^2-1} }
 &=& \frac{ \sqrt{\gamma^2-1}  (\hat{s}\cdot \hat{v}) - \sqrt{\gamma_0^2-1} (\hat{s}\cdot \hat{v}) }{ \sqrt{\gamma_0^2-1}}  \nonumber \\ 
&\approx & -  \frac{e \gamma_0}{2 m (\gamma^2_0-1)}  \kappa A_r \left[  A_r  \sin^2(\omega_r T) +  4 x_e \sin^2 \left(\frac{\omega_r T}{2} \right) \right]   \sin({\omega}_T T) \nonumber \\ &\approx &  
- 1.6 \cdot 10^{-6} \left[\sin^2 \left(\omega_r T\right) + \left( \frac{x_e}{5.625 {\rm mm}} \right)  \sin^2\left(\frac{\omega_r T}{2}\right)  \right]   \sin({\omega}_T T) \, . \nonumber \\
\label{radialresidualR}
\end{eqnarray}
where $x_e$ is typically comparable to the half radius of the storage volume.

\begin{figure}
  \begin{tabular}{p{0.45\textwidth} p{0.45\textwidth} p{0.45\textwidth} p{0.45\textwidth}}
  % p{0.3\textwidth}}
 \vspace{0pt}  \includegraphics[width=0.45 \textwidth]{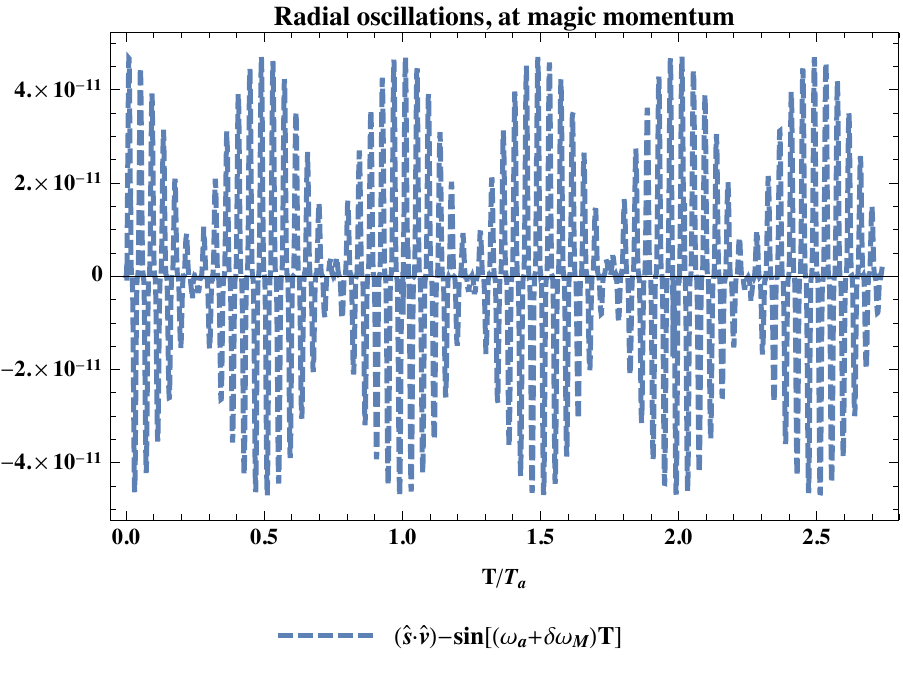} &
  \vspace{0pt}         \includegraphics[width=0.45 \textwidth]{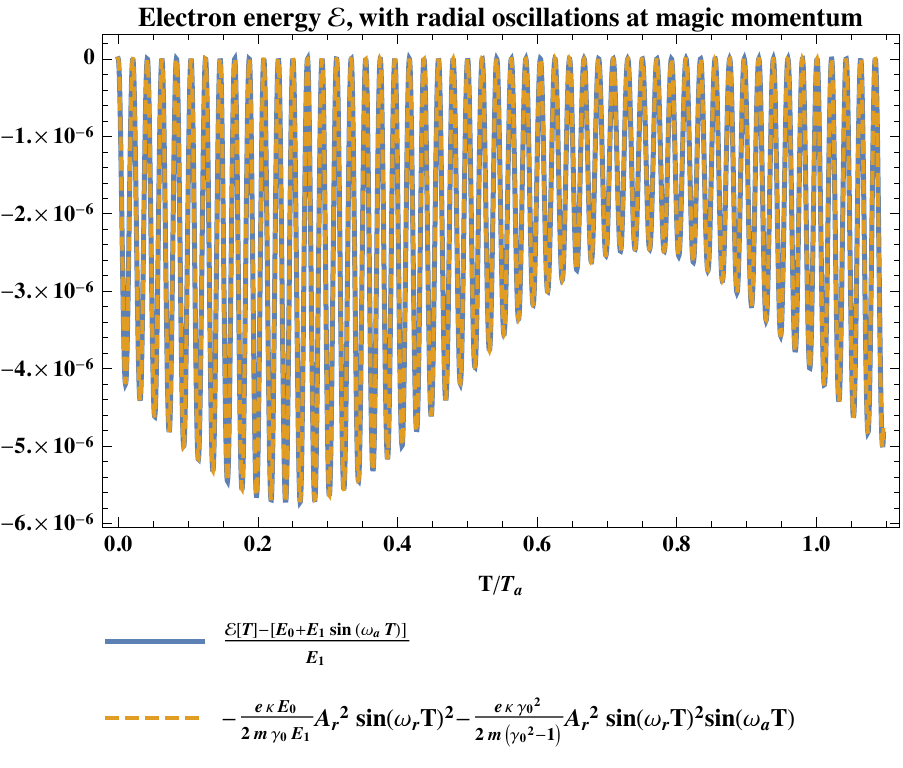} & \\
   \vspace{0pt}         \includegraphics[width=0.45 \textwidth]{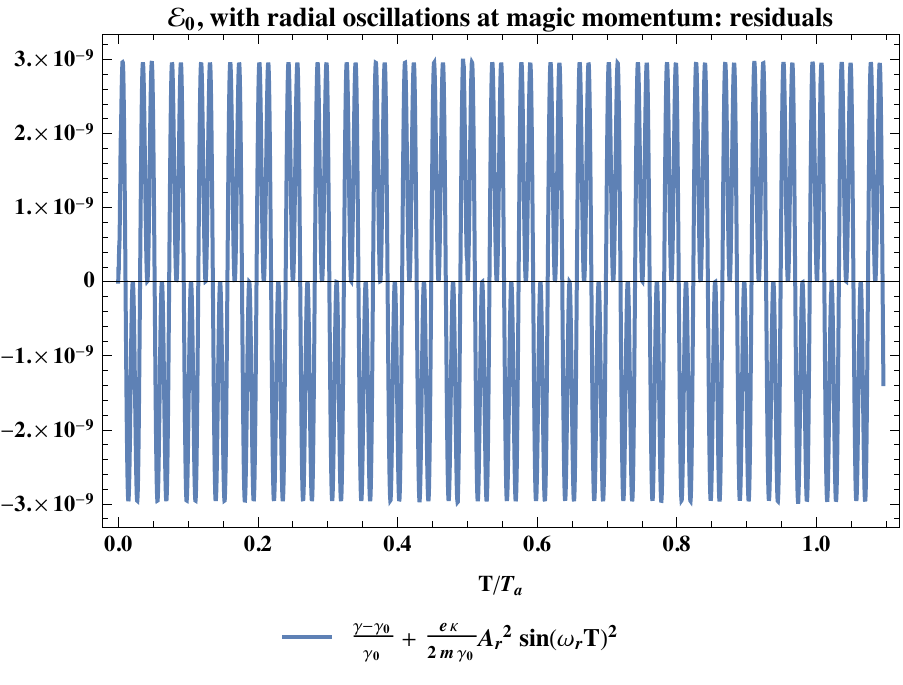} &
      \vspace{0pt}     \hspace{-16pt}    \includegraphics[width=0.52 \textwidth]{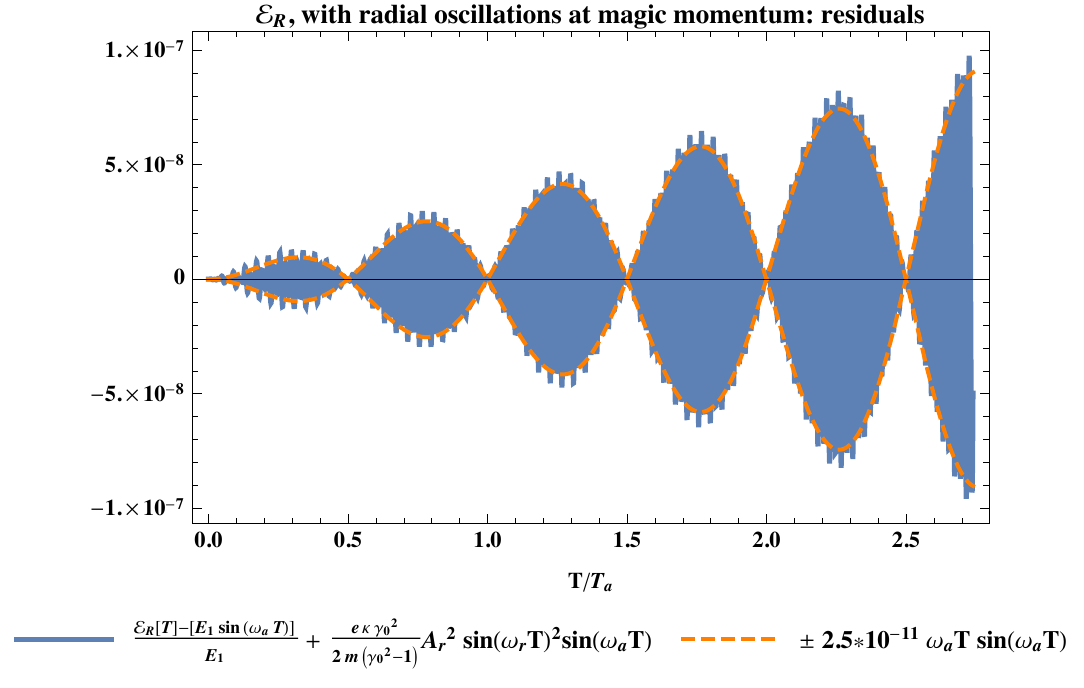} &
%    \vspace{-1pt}        \includegraphics[width=0.3 \textwidth]{Vdependence.pdf} 
    \end{tabular}  %  
%
%	\includegraphics[height=6.7cm]{radialsv} 
%	\includegraphics[height=7.2cm]{energyrad} 
%	\includegraphics[height=5cm]{energyradresidual2} 
%	\includegraphics[height=5cm]{energyradresidual} 
%	\,

	\caption{We have  solved numerically the equations of motion, with a radial displacement, and then evaluated $\hat{s}\cdot \hat{v}=(\Sigma \cdot v)/(\gamma |v| |S|)$ and the electron energy ${\cal E}$ on the solution. We have used here $x_e=0$. The upper left plot shows $\hat{s}\cdot \hat{v}$, compared to a simple sinusoidal behavior: the best fit shows a tiny offset $\delta\omega_M\approx 10^{-11} \omega_a$. The upper right plot shows the energy ${\cal E}$, compared to a simple sinusoidal behavior $E_0+E_1 \sin(\omega_a T)$, where $E_0$ and $E_1$ are constants, and its residual. The bottom plots show the residuals, for ${\cal E}_0$ and ${\cal E}_R$. A very small higher order residual is present for ${\cal E}_R$, which grows in time, signaling a tiny correction $\delta\omega$ to the frequency, since $\cos((\omega_a+\delta\omega) T) \approx \cos(\omega_a T)- T \delta\omega \sin(\omega_a t)$. Here $T_a\equiv 2\pi/\omega_a$.}
\label{energyrad}
%	}
\end{figure}

Also in this case our numerical results can be seen in  fig.~\ref{energyrad}, for the case $x_e=0$. We confirm the existence of such effects, modulo a small frequency shift $\omega_r\rightarrow \omega_r+\delta\omega_r$, where $\delta\omega_r\approx -8\cdot 10^{-5} \omega_r$.

%
%
%\begin{figure}
%  \begin{tabular}{p{0.45\textwidth} p{0.45\textwidth}}
%  % p{0.3\textwidth}}
% \vspace{0pt}  \includegraphics[width=0.44 \textwidth]{energyvert} &
%  \vspace{2pt}         \includegraphics[width=0.45 \textwidth]{energyvertresidual} &
%%    \vspace{-1pt}        \includegraphics[width=0.3 \textwidth]{Vdependence.pdf} 
%    \end{tabular}  %  
%%	\includegraphics[height=6.3cm]{energyvert} 
%%	\includegraphics[height=5.3cm]{energyvertresidual} 
%%	\,
%\label{pitch}.
%	\caption{We have  solved numerically the equations of motion and then evaluated $\hat{s}\cdot \hat{v}=(\Sigma \cdot v)/(\gamma |v| |S|)$ on the solution. The left plot shows that a better fit is achieved by adding the subleading correction $\delta\omega_{SL}$ of Eq.~(\ref{SL}) to the traditional ``pitch correction'', contained in $\tilde{\omega}_a$. Here $T_a\equiv 2\pi/\omega_a$. The residual is well fit (right plot) by the functional form $\epsilon \sin(2\omega_z T) \cos((\tilde{\omega}+\delta\omega_{SL}))$, with $\epsilon=2.436\times10^{-8}$.}
%%	}
%\end{figure}
%
%

Summing all contributions, in the particular case $\varphi_{0, r}=\varphi_{0, z}=0$, we get
\begin{eqnarray}
\frac{\Delta {\cal E}_0}{{\cal E}_0}  &=&  \frac{e \kappa}{2 m \gamma_0}  \Bigg[  A_z^2 \sin ((\omega_z+\omega_r) T ) \sin ((\omega_z-\omega_r) T ) + (A_z^2-A_r^2) \sin^2(\omega_r T)  +  \nonumber \\
&-& 4 x_e A_r  \sin^2 \left(\frac{\omega_r T}{2} \right)  \Bigg] \, ,
\end{eqnarray}
and
\begin{eqnarray}
\frac{\Delta{\cal E}_R}{ {\cal E}_1 |S| \sqrt{\gamma_0^2-1} }  &=&  \frac{e \gamma_0}{2 m (1+\gamma^2_0)} \Bigg[ A_z^2 \sin ((\omega_z+\omega_r) T ) \sin ((\omega_z-\omega_r) T ) + (A_z^2-A_r^2) \sin^2(\omega_r T)  +   \nonumber \\
 &-& 4 x_e A_r  \sin^2 \left(\frac{\omega_r T}{2} \right)  \Bigg] \sin (\omega_a T) \, .
\end{eqnarray}

Such corrections due to a non-constant $\gamma$ should be taken into account in a fit and could significantly affect the $\chi^2$ determination. Nonetheless, since they oscillate rapidly, one expects them not to contaminate much the measurement of $a$. 
However one should also analyze if, in a given the experimental setup, other typical frequencies might play a role in Eq.~(\ref{Eelectron}). If they are close to $\omega_a$, as in the case of the so-called Coherent Betatron Oscillations (CBO), one should check if this could introduce possible systematics, which should be evaluated properly in an experiment. The CBO oscillations are low-frequency oscillations of the beam, in which the width and centroid of the beam oscillate at frequency
$$\omega_{\rm CBO} \approx 2\pi \times 0.476 \,  {\rm MHz} \, .$$

If we assume an amplitude for such oscillations similar to the previous amplitudes $A_z$ or $A_r$, and a sinusoidal behavior as in the previous case, the oscillations in ${\cal E}$, due to $\dot{\gamma}\neq 0$, might be potentially relevant. Such a detailed analysis goes beyond the scope of the present paper and we leave this for future study.

\begin{figure}
  \begin{tabular}{p{0.45\textwidth} p{0.45\textwidth} p{0.45\textwidth} p{0.45\textwidth}}
  % p{0.3\textwidth}}
 \vspace{0pt}  \includegraphics[width=0.45 \textwidth]{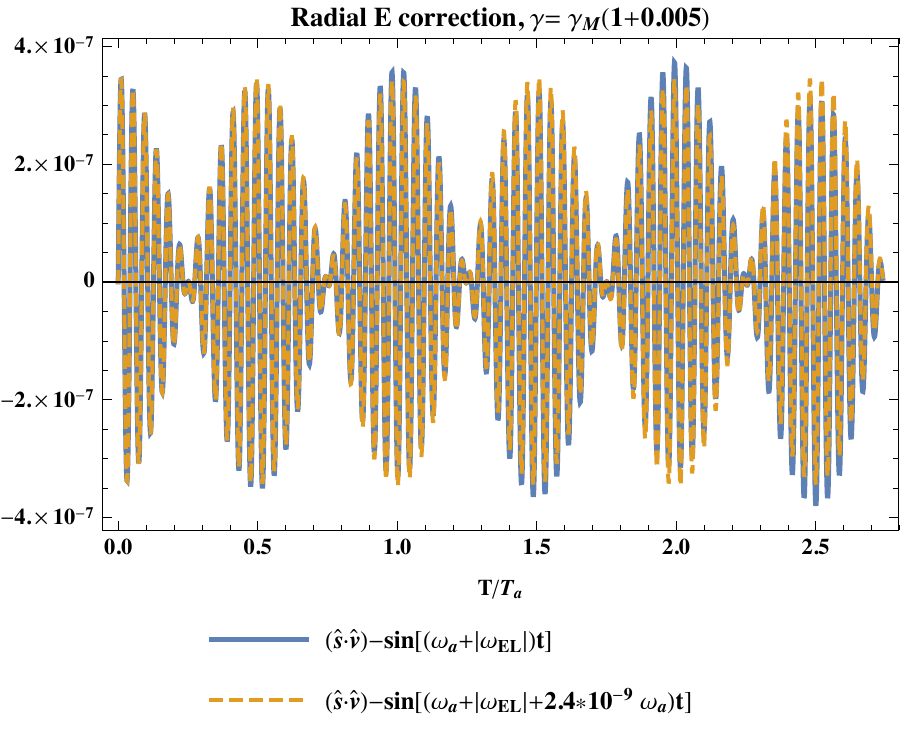} &
  \vspace{0pt}         \includegraphics[width=0.45 \textwidth]{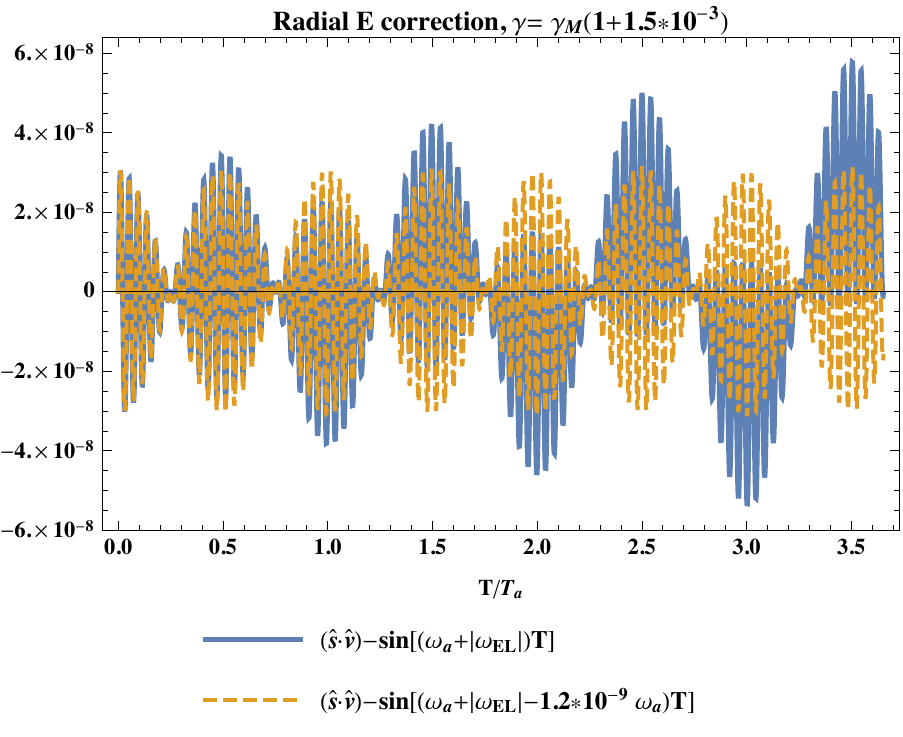} & \\
   \vspace{0pt}         \includegraphics[width=0.45 \textwidth]{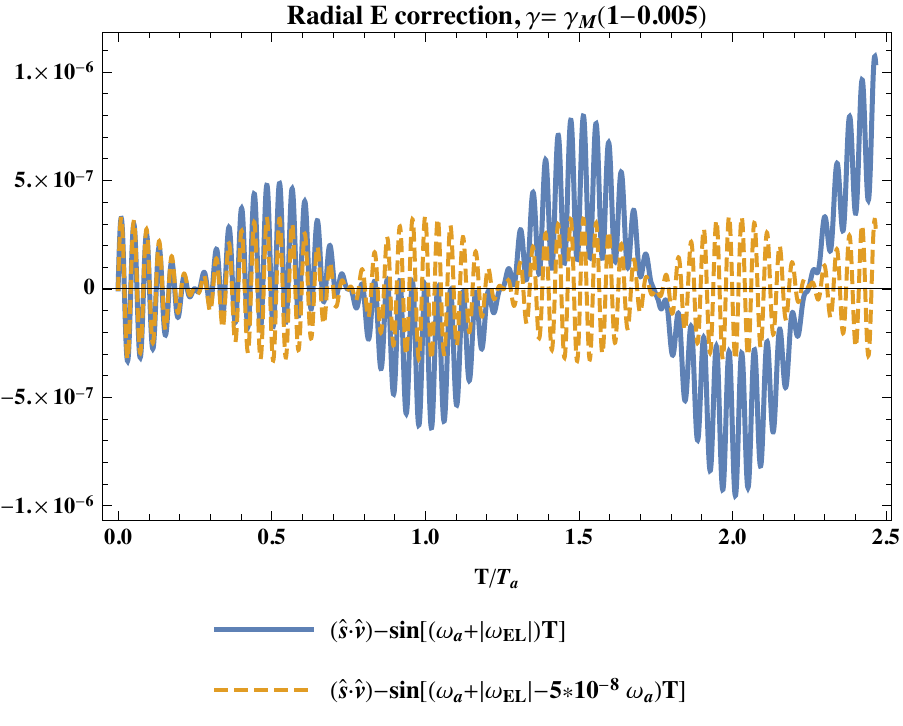} &
      \vspace{0pt}       \includegraphics[width=0.45 \textwidth]{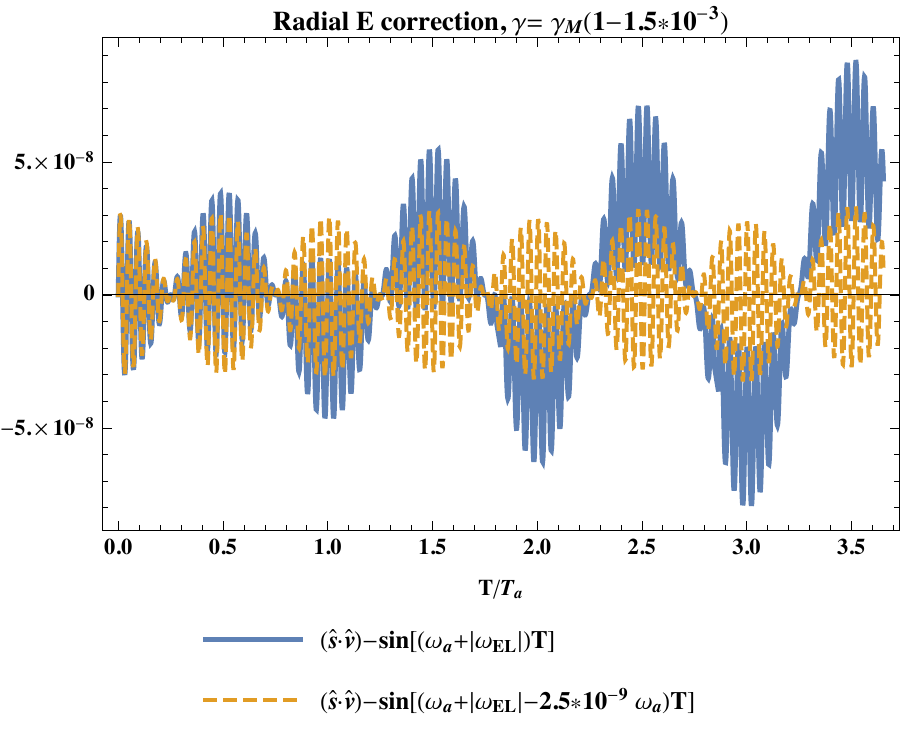} &
%    \vspace{-1pt}        \includegraphics[width=0.3 \textwidth]{Vdependence.pdf} 
    \end{tabular}  %  
%
%	\includegraphics[height=6.7cm]{radialsv} 
%	\includegraphics[height=7.2cm]{energyrad} 
%	\includegraphics[height=5cm]{energyradresidual2} 
%	\includegraphics[height=5cm]{energyradresidual} 
%	\,
%\label{energyrad}.
%	\caption{We have  solved numerically the equations of motion and then evaluated $\omega\equiv \omega_a \left( 1- \frac{(\hat B\cdot \Sigma)(\hat \Sigma \cdot \hat B))}{(\hat \Sigma\cdot v)}\right)^{1/2}$, from Eq.~(\ref{secondderivative}) and $|\tilde{\bfomega}_a|$ (which contains the traditional ``pitch'' correction) from Eq.~(\ref{omegaPsquared}), on the solution. The plot  shows that taking a time average over the fast oscillations (green line) we get precisely the subleading correction $\delta\omega_{SL}$ of Eq.~(\ref{SL}). Here $T_a\equiv 2\pi/\omega_a$ and we used $s^z_0=0$.}
%%	}
%\end{figure}
%
%
%
%\begin{figure}
%	\includegraphics[height=7.3cm]{OmegaradialE} 
%	\,
	\caption{We have  solved numerically the equations of motion on equilibrium orbits with $\gamma\neq \gamma_M$ and so with $x_e\neq0$. We have then evaluated $\hat{s}\cdot \hat{v}=(\Sigma \cdot v)/(\gamma |v| |S|)$ on the solution and compared with a sinusoidal behavior. The plot shows that a better fit is achieved by adding the subleading correction $\delta\omega_{EL,2}$ to the radial electric field correction of Eq.~(\ref{electricR}). The correction turns out to vary with $\gamma_0$ between $10^{-9} \omega_a$ and ${\rm few} \times 10^{-8} \omega_a$ and in a non-symmetric way for values of $\gamma_0$ around the magic momentum $\gamma_M$. }
\label{omegaradialE}
%	}
\end{figure}

Finally, as it is well known, the existence of a radial displacement $x_e$ induces also a radial electric field correction to the precession frequency. This is usually taken into account by considering $\tilde{\bfomega}_a=\bfomega_{\bf v}-\bfomega_{\bf s}$, as in Eq.~(\ref{omegawithpitch}), with the addition of electric fields\footnote{See Eqs.~(11.168) and (11.170), in \cite{Jackson:1998nia}}. This leads to a term $\bfomega_{\rm EL}$ of the form
\begin{eqnarray}
\bfomega_{\rm EL}= \frac{e}{m} \left[ \left(a-\frac{1}{\gamma^2-1}\right) {\bf v}\times {\bf E} \right] \, ,
\end{eqnarray}
which vanishes at the magic momentum $\gamma_R$. In the BNL experiment~\cite{Bennett:2006fi} the distribution of momenta has a spread around $\gamma_R$: the r.m.s. of the energy distribution is of about $0.15\%$, while the maximal possible deviations from the central value is of about $0.5\%$. This induces a correction, which is usually estimated, after averaging the radial oscillations over time, as~\cite{Bennett:2006fi}
\begin{eqnarray}
\frac{\langle |\bfomega_{\rm EL}| \rangle}{\omega_a} \approx -2n (1-n) v_0^2 \frac{x_e^2}{R_0^2} \, . \label{electricR}
\end{eqnarray}
We confirm numerically the existence of such a correction in fig.~\ref{omegaradialE}, but we find also a small subleading correction $\delta\omega_{EL, 2}$ between $10^{-9} \omega_a$ and ${\rm few} \times 10^{-8} \omega_a$, depending on the value of $\gamma_0$, taken within $0.15\%-0.5\%$ around the magic momentum. Moreover the subleading correction turns out to be asymmetric whether $\gamma_0$ is larger or smaller than the magic momentum $\gamma_M$. Such a numerical subleading correction should be then averaged over the distribution of momenta in a realistic experiment and could lead to a non-trivial overall correction, presumably leading to a systematic error of at most $\Delta\omega/\omega_a\approx 10^{-8}-10^{-9}$.

The existence of such subleading corrections could be expected, as we have learned from the case of vertical oscillations that the approach of considering the second derivative of the energy ${\cal E}$ and taking a time average better reproduces the behavior of the system. In presence of electric fields this is given by Eq.~(\ref{lengthy}), which could be averaged over time. 
At the central radius of the experiment $R_0$ the electric field is vanishing and so all corrections (except for the ``pitch'' correction, if there are vertical oscillations) vanish. As soon as the muons are displaced from $R_0$ there will be several corrections, which, as we have already noted, do not vanish at $\gamma=\gamma_M$.

\section{Conclusions}

We have derived a full set of equations for  the energy ${\cal E}$ of the decay electrons, in a muon $g-2$ experiment. We have considered the equations for velocity and spin of the muon in an electromagnetic field, in a generic metric and seen by a generic observer. The energy ${\cal E}$ is modulated mainly by the scalar product of the observed spin $\Sigma$ and the observed  velocity $v$.  As it is well known, in a cyclotron in the horizontal plane and in flat spacetime, with a constant strong magnetic field of amplitude $|B|$ along the vertical axis, we have that ${\cal E}$ oscillates in time, at the standard anomalous frequency $\omega_a=ea |B|/m$. We have studied then several corrections to this leading behavior.

First, we have analyzed the frequency shift $\delta\omega_P$, due to the so-called ``pitch correction'', that arises when the muon trajectory is not purely orthogonal to the magnetic field. In a realistic cyclotron, indeed, muons perform small oscillations of amplitude $A_z$ and angular frequency $\omega_z$, along the vertical direction of the cyclotron $\hat{z}$. The commonly used estimate of the pitch correction is determined by the time average of the velocity $\langle (v^z)^2 \rangle /2$, leading to $\delta \omega_P/\omega_a= A_z^2 \omega_z^2/4\approx 10^{-6}$. We find a good numerical agreement with such an estimate, but we find a subleading correction $\delta \omega_{SL}/\omega_a \approx {\cal O} (10^{-9})$, which might be relevant in future experiments. We demonstrate the existence of such a subleading correction analytically, by defining the  frequency through the second derivative of the observed energy ${\cal E}$ and showing that it depends on the time evolution of the spin $\Sigma^z$. After averaging over time, we find good agreement between the numerical and analytical results.

Then, we have analyzed another well-known correction due to radial electric fields: such a correction is also commonly estimated with the term $\frac{e}{m} \left[ \left(a-\frac{1}{\gamma^2-1}\right) {\bf v}\times {\bf E} \right] $ , which vanishes at the so-called magic momentum. Also in this case we find numerically subleading corrections $\delta \omega_{SL}/\omega_a \approx {\rm few}\times 10^{-8} \sim 10^{-9}$, which should be related to the existence of many correction terms in the second derivative of ${\cal E}$, which do not all vanish at the magic momentum.

Then we have analyzed rapidly oscillating terms in ${\cal E}$, due to the time dependence of the muon $\gamma$ factor, in the presence of electric fields. In a cyclotron such corrections  arise due to oscillations of the muons along both the radial and vertical direction and are in principle non-negligible, being of relative size $\Delta {\cal E}/{\cal E}\approx 10^{-6}$ in a realistic experiment. It remains to be checked whether such terms, which oscillate at high frequency, might have an impact in a fit of a realistic experiment. It also remains to extend the analysis including the so-called lower frequency Coherent Betatron Oscillations, which are potentially more dangerous.

Finally, we have analyzed the extra terms introduced by the presence of a non-trivial metric, finding that they depend on the gravitational ``force'' $f_G$.
We have considered  two specific cases: a Schwarzschild metric and a rotating metric, representing the Earth's gravity and rotation respectively.

For the Schwarzschild case the effects are tiny: the leading term in $f_G$ is due to the effect of the standard surface gravity, $g_T=9.8 \, {\rm m}/{\rm s}^2$, on the spin vector, leading to a relative correction of order $g_T/(\gamma \omega_a) \approx 8\cdot 10^{-16}$, at the so-called magic momentum.

The leading effects due to a non-trivial metric are instead due to rotation. We worked at linear order in the angular frequency $\omega_T$, finding that $f_G$ contains a Coriolis force, which amounts to an additional frequency vector ${\bf \Omega}_T/\gamma$, with $|\Omega_T|=\omega_T$, aligned with the Earth's rotation. This corresponds to an angular frequency of $\Delta\omega \approx 2\pi \times 4\times 10^{-7}$ Hz, which represents a relative correction of about $\Delta\omega/\omega_a\approx 1.2\times 10^{-12}$. This is very small for muon $g-2$ experiments, but it could be more important for {\it electron} $g-2$ experiments. In the latter case the Coriolis frequency vector is just ${\bf \Omega}_T$, to be compared with the Larmor frequency of about $\omega_L=100 \, {\rm GHz}$, for an electron in a Penning trap experiment~\cite{Hanneke:2008tm,Sturm:2013upa}. This gives a relative effect $\Delta\omega/\omega_L\approx 7\times 10^{-16}$ to be compared with the experimental relative error on $g$, at present of order ${\cal O}(10^{-13})$ and forecasted to reach ${\cal O}(10^{-14})$ ~\cite{Gabrielse:2019cgf} within one or two years.

\acknowledgments
We thank Massimo Passera, Gianguido dall'Agata and Stefano Laporta for useful discussions. We also thank James P.~Miller and B.~Lee Roberts and Marco Incagli, for useful correspondence. A.N. was supported by the grants FPA2016-76005-C2-2-P, MDM-2014-0369 of ICCUB (Unidad de Excelencia Maria de Maeztu), AGAUR 2014-
SGR-1474, SGR-2017-754. DB acknowledges partial financial support by ASI Grant No. 2016-24-H.0.
A.N. is grateful  to the Physics Department of the University of Padova for the hospitality and was supported by the ``Visiting Scientist'' program of the University of Padova during the initial stages of this project.
\appendix

\section{Appendix A}\label{Sec:calS}

In this Appendix we want to examine the precession rate of the spin $S$ w.r.t. the adapted frame $\{e_{\rm (M)}^{\hat a}\}$ orthogonal to $U=e^{\rm (M)}_{\hat 0}$. 
Following \cite{Felice:2010cra} (see also \cite{Jantzen:1992rg}), this is equivalent to the orientation w.r.t the observer $u$, described with the adapted frame $\{e^{\hat \mu}\}$,  of a boosted spin vector $\calS$, which is obtained with the following relation%Let us define the 
\be \label{calS}
\calS=\Sigma - \frac{\gamma}{1+\gamma} (v \cdot \Sigma) v\;.
\ee
We note immediately that Eq.~(\ref{calS}) is a generalisation (in GR) of the inverse function of Eq.~(\ref{Slab}), {\it e.g.} see  \cite{Jackson:1998nia}.
 The boosted spin vector $\calS$ has the following properties
\be
u\cdot \calS=0\;, \quad \quad   P(u) \calS=\calS \, , % \quad \quad P(u,U)\calS =\Sigma
 \quad \quad v\cdot \Sigma = \gamma (v \cdot \calS)\;.
\ee
Using
\be
\Sigma = S - (v \cdot \Sigma) u =S - \gamma (v \cdot \calS) u \, ,
\ee
and Eq. (\ref{calS}), we can write the map between $\calS$ and $S$, {\it i.e.}
\be \label{calS2}
S=\calS + \frac{\gamma}{1+\gamma} (v \cdot \calS) (U+u)\;.
\ee
Differentiating both sides of the above relation  along $U$ and taking into account Eq.~(\ref{EqSpin}) we are able to write the the following equation
\bea\label{DcalS/dtau}
{D\calS\over d\tau}  &=&\frac{e(1+a)}{m} F * \left[\calS + \CcalS (U+u)\right]-a\gamma  \left[(f_{\rm EM} \cdot \calS) - \CcalS (v \cdot f_{\rm EM})\right]U -  (U+u) {d \CcalS  \over d \tau}\nonumber\\
&&+\left({DU\over d\tau} - f_G \right) \CcalS \;,
\eea
where we have defined
\[\CcalS \equiv  \frac{\gamma}{1+\gamma} (v \cdot \calS)\, , \] 
and used the following relations
\[u \cdot {DU\over d\tau} = -\gamma (v \cdot f_{\rm EM}) \quad \quad {\rm and}  \quad \quad P(u){DU\over d\tau} =\gamma  f_{\rm EM}\;. \]
Now, projecting Eq. (\ref{DcalS/dtau}) with $P(u)$, we have
\bea \label{PuDcalS/dtau}
P(u){D\calS\over d\tau} &=& {D_{({\rm fw},U,u)}\calS\over d\tau} =\frac{e(1+a)}{m} \left[P(u)F * \calS \right]  + \frac{e(1+a)}{m}  \CcalS \left[P(u) F *  (U+u)\right] \nonumber\\
&&-a\gamma^2  \left[(f_{\rm EM} \cdot \calS) - \CcalS (v \cdot f_{\rm EM})\right]v- \gamma {d \CcalS  \over d \tau} v +  \CcalS (- \gamma   f_{\rm EM} + f_G )\;;
\eea
instead, projecting Eq.~(\ref{DcalS/dtau}) along $u$, we find
\bea \label{uDcalS/dtau}
u \cdot {D\calS\over d\tau} &=& - {D u\over d\tau} \cdot \calS=f_G \cdot \calS=\frac{e(1+a)}{m} \left[u* F * \calS \right] + \frac{e(1+a)}{m}  \CcalS \left(u* F *  U\right)  \nonumber\\
&& + a\gamma^2  \left[(f_{\rm EM} \cdot \calS) - \CcalS (v \cdot f_{\rm EM})\right] +(1+\gamma) {d \CcalS  \over d \tau}  + \gamma \CcalS (v \cdot f_{\rm EM}  )\;.
\eea
In this way, using Eq.~(\ref{uDcalS/dtau}) we can remove ${d \CcalS  / d \tau}$ within (\ref{PuDcalS/dtau}) and we obtain the final result
 \bea
 {D_{({\rm fw},U,u)}\calS\over d\tau} &=&\frac{e(1+a)}{m} \left[P(u)F * \calS \right]  + \frac{e(1+a)}{m}  \CcalS \left[P(u) F *  (U+u)\right] \nonumber\\
&&-a\gamma^2  \left[(f_{\rm EM} \cdot \calS) - \CcalS (v \cdot f_{\rm EM})\right]v  +  \CcalS (- \gamma   f_{\rm EM} + f_G )\nonumber\\
&& - {\gamma \over \gamma +1}\Bigg\{ - \frac{e(1+a)}{m} \left[u* F * \calS \right] - \frac{e(1+a)}{m}  \CcalS \left(u* F *  U\right)  \nonumber\\
&&+  \left[- a\gamma^2  f_{\rm EM} + f_G \right]   \cdot \calS  - \gamma \CcalS (v \cdot f_{\rm EM}) \left[-a\gamma +1 \right] \Bigg\} \, ,
%- \gamma {d \CcalS  \over d \tau} v +  \CcalS (- \gamma   f_{\rm EM} + f_G )
 \eea
 or, equivalently,
\be 
\left({d \calS \over d T }\right)^{\hat a} = \left[\Omega_\calS \times_u \calS\right]^{\hat a}\;,
\ee
where
\bea
\Omega_\calS &=& {1 \over \gamma+1}  v \times_u ( f_G -\gamma f_{\rm EM}) + {e(1+a)\over m}\left\{ v \times_u \left[ E +{1 \over \gamma+1}(v \times_u B) \right] - {B\over \gamma}\right\}\nonumber\\
&&- (\zeta_{({\rm fw)}}+\zeta_{({\rm sc})})\;.
\eea

%\hat \times_u \Bigg\{ \frac{(1+a)e}{m \gamma}\left[(v \cdot \hat\Sigma)E+\left( \hat \Sigma  \times_u  B \right)\right]+ \frac{1}{\gamma}(v \cdot \hat\Sigma)f_G +\gamma a \left[(f_{\rm EM} \cdot\hat\Sigma) - (v\cdot\hat\Sigma) (f_{\rm EM}\cdot v)\right]v\Bigg\} - (\zeta_{({\rm fw)}}+\zeta_{({\rm sc})})\;.

\bibliographystyle{JHEP}

\bibliography{CORIOLIS}

\end{document}